\documentclass[journal=jacsat,manuscript=article]{achemso}

\usepackage{amsmath}
\usepackage{amsmath,amssymb}
\usepackage{graphicx}
\usepackage{caption}
\usepackage{color}
\usepackage{dcolumn}
\usepackage{bm}
\usepackage{float}
\usepackage{subfig}
\usepackage{url}
\usepackage{xr}
\author{Marco Pezzella} \affiliation{Department of Chemistry,
  University of Basel, Klingelbergstrasse 80, CH-4056 Basel,
  Switzerland}
  
\author{Krystel El Hage} \affiliation{Department of Chemistry,
  University of Basel, Klingelbergstrasse 80, CH-4056 Basel,
  Switzerland} \altaffiliation{SABNP, Univ. Evry, INSERM U1204,
  Universit\'e Paris-Saclay, 91025 Evry, France}

\author{Michiel J.M. Niesen} \affiliation{Department of Chemistry,
    MIT, USA}

\author{Sucheol Shin} \affiliation{Department of Chemistry, University
  of Texas at Austin, USA}

\author{Adam P. Willard}\email{awillard@mit.edu}
\affiliation{Department of Chemistry,  MIT, USA} 
  
\author{Markus Meuwly}\email{m.meuwly@unibas.ch}
\affiliation{Department of Chemistry, University of Basel,
  Klingelbergstrasse 80, CH-4056 Basel, Switzerland}

\author{Martin Karplus}\email{marci@tammy.harvard.edu}
\affiliation{Department of Chemistry,
  Harvard University, USA}
\affiliation{Laboratoire de Chimie Biophysique, ISIS, Universit\'{e} Louis
  Pasteur, 67000 Strasbourg, France}

\title{Water Dynamics Around Proteins: T- and R-States of Hemoglobin
  and Melittin}

\begin{document}

\date{\today}

\begin{abstract}
The water dynamics, as characterized by the local hydrophobicity (LH),
is investigated for tetrameric hemoglobin and dimeric melittin. For
the T$_0$ to R$_0$ transition in Hb it is found that LH provides
additional molecular-level insight into the Perutz mechanism, i.e.,
the breaking and formation of salt bridges at the $\alpha_1 / \beta_2$
and $\alpha_2 / \beta_1$ interface is accompanied by changes in LH.
For Hb in cubic water boxes with 90 \AA\/ and 120 \AA\/ edge length it
is observed that following a decrease in LH as a consequence of
reduced water density or change of water orientation at the
protein/water interface the $\alpha / \beta$ interfaces are
destabilized; this is a hallmark of the Perutz stereochemical model
for the T to R transition in Hb. The present work thus provides a
dynamical view of the classical structural model relevant to the
molecular foundations of Hb function. For dimeric melittin, earlier
results by Cheng and Rossky [Nature, 1998, 392, 696–699] are confirmed
and interpreted on the basis of LH from simulations in which the
protein structure is frozen. For the flexible melittin dimer the
changes in the local hydration can be as much as 30 \% than for the
rigid dimer, reflecting the fact that protein and water dynamics are
coupled.
\end{abstract}

\section{Introduction}
Hemoglobin is one of the most widely studied proteins due to its
essential role in transporting oxygen from the lungs to the
tissues. The two most important structural states of this protein are
the deoxy structure (T$_0$), which is stable when no ligand is bound
to the heme-iron, and the oxy structure (R$_4$), which is stable when
each of the four heme groups have a ligand, such as oxygen, bound to
them. The state with the quaternary structure of R$_4$, but with no
heme-bound ligands is the R$_0$ state. Despite strong experimental
evidence that T$_0$ is significantly more stable than R$_0$, with an
equilibrium constant of $K_{\frac{T_0}{R_0}}=6.7 \times 10^5$,
\cite{edelstein:1971} molecular dynamics (MD) simulations appear to
indicate that the R$_0$ state is more stable.  Specifically,
simulations have found that when hemoglobin is initialized in the
T$_0$ state it undergoes a spontaneous transition into the R$_0$ state
on sub-$\mu$s time scales.\cite{Hub_2010,yusuff:2012} Understanding
the origins of this discrepancy between the measured and simulated
relative stabilities of the R$_0$ and T$_0$ states is essential to
establishing the reliability of simulation-based studies of Hemoglobin
and other large biomolecules. \\

\noindent
In a recent simulation study, it was found that the T$_0$
$\rightarrow$ R$_0$ transition rate depends sensitively on the size of
the simulation cell.\cite{MM.hb:2018} Specifically, simulations of
hemoglobin initialized in the T$_0$ state and placed in a periodically
replicated cubic solvent box with side length of 75 \AA, 90 \AA, and
120 \AA, underwent transition towards the R-state structure after 130
ns, 480 ns, and 630 ns, respectively. Furthermore, in a simulation box
with side-length of 150 \AA, hemoglobin remained in the T$_0$ state
for the entirety of a 1.2$\mu$s simulation. The extrapolated trend in
these findings implies that T$_0$ is the thermodynamically stable
state in this largest simulation cell for which the diffusional
dynamics of the environment are correctly captured. The results also
suggested that such a large box is required for the hydrophobic
effect, which stabilizes the T$_0$ tetramer, to be manifested. While
the statistical significance of this conclusion has been a topic of
recent discussion in the literature,\cite{degroot:2019,MM.hb:2019} the
dynamic stability of the T$_0$ state exhibits a clear systematic
dependence on the size of the solvent box. Further analysis is
required to provide conclusive evidence of the role of the hydrophobic
effect and to reveal the mechanistic origin of the dependence of the
thermodynamic stability of the T$_0$ state on the simulation box
size. In this study we specifically address the role of system size
variations in solvent dynamics.\\

\noindent
The present work addresses the system size question by analyzing the
molecular structure of the hydration layers surrounding tetrameric
hemoglobin (Hb) and dimeric melittin. The particular focus is on
whether there are characteristic changes in local hydration that
accompany global transitions involving reorientation of the subunits -
i.e. the decay of the T-state for Hb and the reconfiguration of the
helices in melittin - and whether and how these changes are effected
by the size of the solvent box. Extending the study to the melittin
dimer, which is much smaller than Hb, provides information about the
generality of this analysis. In addition, melittin was also studied
previously as an example for hydrophobic hydration.\cite{rossky:1998}
Melittin is a small, 26-amino acid protein found in honeybee venom
that crystallizes as a tetramer, consisting of two dimers, related by
a two-fold symmetry axis.\cite{terwilliger1982.1,terwilliger1982.2}
Previous work has characterized the behaviour of the hydrophobic
binding surface of melittin and the solvent exposed surface
residues.\cite{rossky:1998} While these surface residues are
characterized by a well-defined orientation of the water molecules,
water molecules in the hydrophobic regions are more dynamical,
exploring different water configurations. Here, similar simulations
with a frozen melittin dimer in different box sizes are carried out
and analyzed. In addition, the protein is also allowed to move freely
which provides information about the solvent-solute coupling which has
not been considered before.\cite{rossky:1998} \\

\noindent
The analysis is based on a recently developed method of characterizing
the hydrophobicity of a surface based on a statistical analysis of the
configurational geometries of interfacial water
molecules.\cite{shin2018} This method, described in more detail in the
``Analysis of Aqueous Interfacial Structure'' section, generates an
order parameter, $\delta \lambda_\mathrm{phob}$, which quantifies the
statistical similarity of sampled water configurations to those that
occur at equilibrium near an ideal hydrophobic surface. When applied
to water configurations sampled from a particular nanoscale region of
a protein surface, $\delta \lambda_\mathrm{phob}$ can be interpreted
as a local measure of hydrophobicity and thus be extended to map the
spatial and temporal variations of a protein's solvation shell.  \\

\noindent
In the following section the simulations and computational methods are
described. Then, in the Results section, analyses and interpretations
of protein hydration structure are presented.  Results for hemoglobin
are described first, including analysis of previous simulations in
different simulation box sizes.  Results for melittin are described
second. Finally, conclusions are drawn.\\

\section{Computational Methods}
\subsection{Molecular Dynamics Simulations}
{\bf Simulations of Hemoglobin (Hb):} The Hb-simulations (for the
sequence see Figure S1) were described
previously,\cite{MM.hb:2018} and only the necessary points without
technical details are reported here. The molecular dynamics
trajectories were run in cubic water boxes with box lengths 90 \AA\/,
120 \AA\/, and 150 \AA\/ which are analyzed in the following. Each
simulation was run for 1 $\mu$s or longer and for selected box sizes,
additional repeat simulations were carried out.\cite{MM.hb:2019} The
trajectories analyzed in the present work are those from
Ref.\cite{MM.hb:2018} and the reader is referred to that manuscript
for additional details on the production runs.\\

\noindent
{\bf Simulations of Melittin:} MD simulations of the melittin dimer
were carried out using CHARMM\cite{CHARMM2009} c45a1 and the CHARMM36
force field.\cite{charmm36ff} The TIP3P water model was used, the same
as that used for Hb. The dimer structure
(PDB:2MLT)\cite{eisenberg:1990} was used as the starting structure. It
was solvated in a cubic water box of length 51.051 \AA\/ (4066 water
molecules) In addition, simulations with a box length of 60 \AA\/ were
performed to assess whether, in analogy to Hb, there were effects of
increased solvent box sizes on the stability, dynamics and water
structuring of melittin dimer. A 16 \AA\/ cut-off was applied with a
Particle Mesh Ewald scheme\cite{spme_darden_1995} and a 1 fs time step
was used in the MD simulations. Although more physically realistic
fixed point charge water models exist (e.g. TIP4P), the use of TIP3P
here is mandatory for consistency because the CHARMM force field was
parametrized with it.  It is certainly of interest to include water
polarization in corresponding simulations,\cite{huang:2016} but such a
study is beyond the scope of the present work.\\

\noindent
The following protocol was used. Two steps of minimization were
performed: 50 steps with the Steepest Descent algorithm, followed by
50 steps with the Newton-Raphson algorithm. The system was then heated
and equilibrated using the velocity Verlet algorithm\cite{velver:1982}
for 25 ps with a Nose Hoover\cite{NoseHoover85} thermostat at 300
K. This was followed by a 100 ns $NVT$ production simulation, for
which coordinates were recorded every 1 ps.  In a first set of
simulations, the protein dimer was fixed and only the solvent water
was allowed to move.  This allows direct comparison with the work of
Cheng and Rossky.\cite{rossky:1998} In a separate set of 100 ns
simulations the protein was allowed to move and only bonds involving
hydrogen atoms molecules were constrained using
SHAKE.\cite{shake-gunsteren}\\

\subsection{Analysis of Aqueous Interfacial Structure}
The hydration structure of the simulated proteins was characterized
following a recently developed computational method.\cite{shin2018}
This method is based on the concept that deformations in water's
collective interfacial molecular structure encode information about
the details of surface-water interactions.\cite{shin.water:2018} These
deformations are quantified in terms of the probability distribution
of molecular configurations, as specified by the three-dimensional
vector, $\vec{\kappa} = (a,\cos{\theta_{\rm OH_1}},\cos{\theta_{\rm
    OH_2}})$, where $a$ is the distance of the oxygen atom position to
the nearest point on the instantaneous water interface, as defined in
Ref.\cite{willard_instantaneous_2010}, and $\theta_{\rm OH_1}$ and
$\theta_{\rm OH_2}$ are the angles between the water OH bonds and the
interface normal.\\

\noindent
Here, this method is used to compute the time dependent quantity,
$\delta \lambda_\mathrm{phob}^{(r)}(t)$, which describes the local
hydrophobicity (LH) of residue $r$, at time $t$. 
More specifically,
$\delta \lambda_\mathrm{phob}^{(r)}(t) =
\lambda_\mathrm{phob}^{(r)}(t) - \langle \lambda_\mathrm{phob}
\rangle_0$, where,
\begin{equation}
    \label{eqn:loglike}
    \lambda_\textrm{phob}^{(r)}(t) = -\frac{1}{{\sum_{a=1}^{N_a(r)}
        N_w(t;a)}}\sum_{a=1}^{N_a(r)} \sum_{i=1}^{N_w(t;a)} \ln{\left[
        \frac{P(\vec{\kappa}^{(i)}(t)|\textrm{phob})}{P(\vec{\kappa}^{(i)}(t)|\textrm{bulk})}
        \right]}.
\end{equation}
Here the summation over $N_a(r)$ is over the atoms in residue $r$ and
the summation over $N_w(t;a)$ is over the water molecules within a
cut-off of 6\AA\/ of atom $a$ at time $t$, and $\vec{\kappa}^{(i)}(t)$
denotes the configuration of the $i$th molecule in this population.
$P(\vec{\kappa}\vert\mathrm{phob})$ is the probability to find
configuration $\vec{\kappa}$ at an ideal hydrophobic surface and
$P(\vec{\kappa} \vert \mathrm{bulk})$ is the probability to find that
same configuration in the isotropic environment of the bulk liquid.
As described in Ref.~\citenum{shin2018}, these reference distributions
were obtained by sampling the orientational distribution of water at
an ideal planar hydrophobic silica surface and the bulk liquid,
respectively. The quantity $\langle \lambda_\mathrm{phob} \rangle_0$
is the equilibrium value of $\lambda_\mathrm{phob}$ for
configurational populations sampled from the ideal hydrophobic
reference system. \\

\noindent
Values of $\delta \lambda_\textrm{phob}^{(r)}$ close to zero indicate
that water near residue $r$ exhibits orientations that correspond to
those found at an ideal hydrophobic surface. Hydrophilic surfaces
interact with interfacial water molecules and lead to configurational
distributions that differ from that of an ideal hydrophobic surface.
These differences are typically reflected as values of $\delta
\lambda_\mathrm{phob}^{(r)} > 0$, with larger differences giving rise
to larger positive deviations in $\delta \lambda_\mathrm{phob}^{(r)}$.
Values of $\delta \lambda_\mathrm{phob}{(r)} \geq 0.5$ are used as
indicative of hydrophilicity.  For the number of unique water
configurations used to compute $\delta \lambda_\mathrm{phob}^{(r)}$
here, fluctuations of $\delta \lambda_\mathrm{phob}^{(r)}$ are
expected to fall within $-0.24 \leq \delta\lambda_\mathrm{phob}^{(r)}
\leq 0.27$ (95\% confidence interval) at the hydrophobic reference
system, making sustained values of $\lambda_\mathrm{phob}^{(r)} \geq
0.5$ highly indicative of local hydrophilicity. The fluctuations in
$\delta \lambda_\textrm{phob}^{(r)}$ as a function of time provide
information about changes in the local solvation environment.\\

\section{Results}

\subsection{Hydration Dynamics around T$_0$- and R$_0$-State Hemoglobin}
\begin{figure}
\centering \includegraphics[scale=0.45]{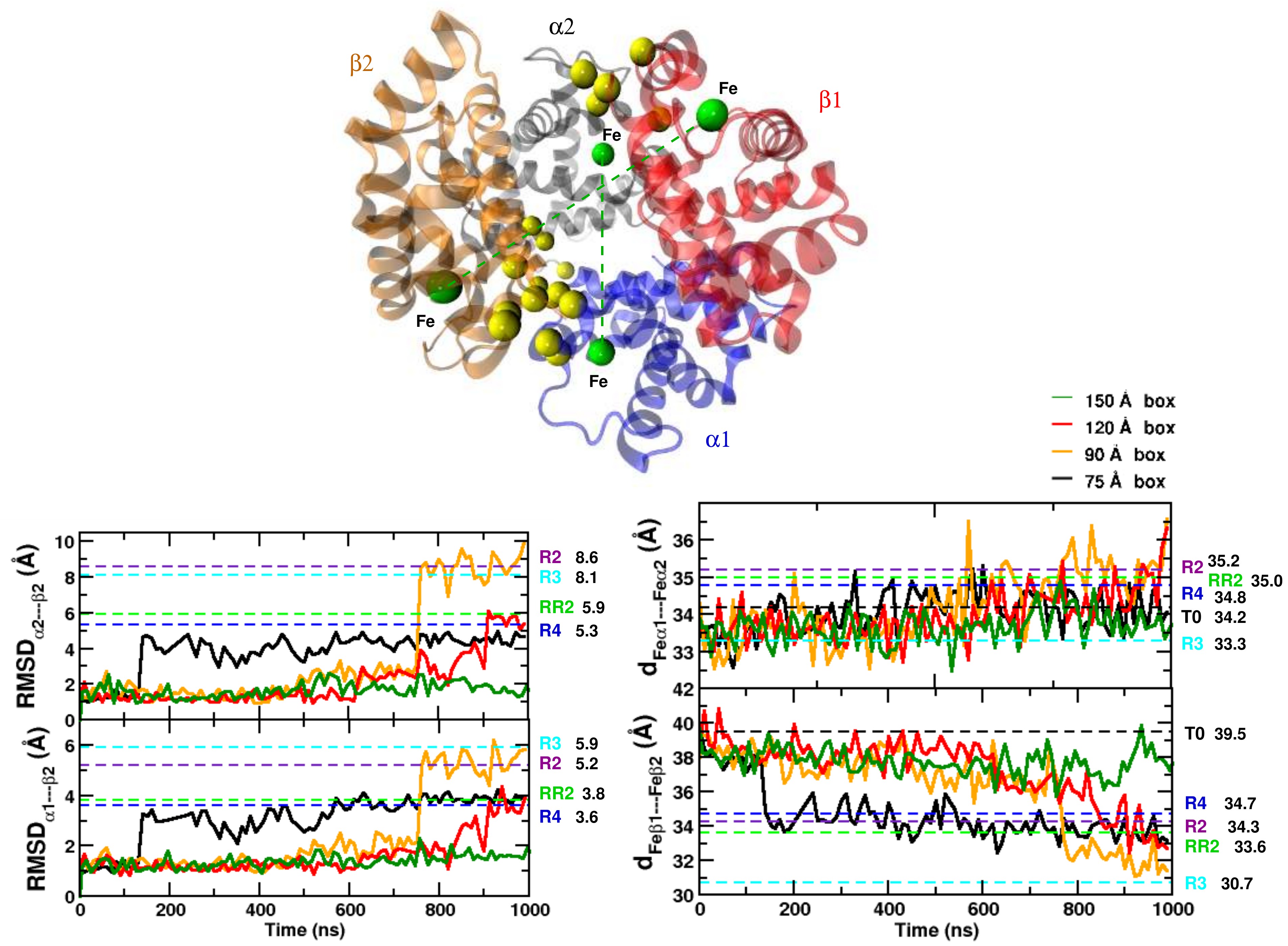}
\caption{Top: Representation of Hb with the C$_\alpha$ atoms of the
  residues relevant to Perutz' stereochemical mechanism shown as
  yellow spheres. The iron atoms (green spheres) are connected with
  green dashed lines, indicating the distances reported. Blue, red,
  grey, and gold ribbon structures for $\alpha_1$, $\beta_1$,
  $\alpha_2$, and $\beta_2$ subunits of Hb.  Bottom: Quaternary
  structure differences of Hb structures found in the simulations,
  based on the structural comparison of the $\alpha_{1} \beta_{1}$ and
  $\alpha_{2} \beta_{2}$ subunits. Black, gold, red, and green traces
  for simulations in the 75 \AA\/, 90 \AA\/, 120 \AA\/, and 150 \AA\/
  boxes, respectively. Left panels: (top) RMSD of the nonsuperimposed
  $\alpha_{2} \beta_{2}$ subunit after superimposing the $\alpha_{1}
  \beta_{1}$ subunit (C$_\alpha$ carbon atoms were used for both
  superposition and RMSD calculation); (bottom) RMSD of the
  nonsuperimposed $\alpha_{1} \beta_{2}$ subunit after superimposing
  the $\alpha_{2} \beta_{1}$ subunit. Right panels: Iron-Iron
  distances between the $\alpha$s and $\beta$s of each
  subunit. Horizontal dashed lines indicate the corresponding values
  from all known Hb structures (T$_0$, R$_2$, RR$_2$, R$_3$ and
  R$_4$).}
\label{fig:struchb}
\end{figure}

Figure \ref{fig:struchb} (top) illustrates the structure of Hb for the
sequences of the $\alpha$ and $\beta$ chains) with the C$_{\alpha}$
atoms of the residues analyzed specifically shown as van der Waals
spheres. The first set of residues we study are the ones in Perutz'
stereochemical model\cite{perutz:1970}, which are involved in the salt
bridges and the $\alpha / \beta$ shearing motion (Table
\ref{tab:tab1}). The shearing motion involves a change in the H-bonds
at the $\alpha_1 / \beta_2$ interface. In the T$_0$ structure, the
side chain of Thr41$\alpha_1$ occupies a notch formed by the main
chain of Val98$\beta_2$ and a hydrogen bond is present between
Tyr42$\alpha_{1}$ and Asp99$\beta_2$. After the transition to the
R$_4$ state, the same notch is occupied instead by Thr38$\alpha_1$ and
the previous hydrogen bond is substituted by one between
Asp94$\alpha_1$ and Asn102$\beta_2$. The same conformational change
occurs at the $\alpha_2/\beta_1$ interface.\\

\noindent
Other R-state structures, including R$_2$, RR$_2$ and R$_3$, also
exist and present intermediate states between T$_0$ and R$_4$.  The
difference between all Hb forms emerges from differences in the
position of the $\beta_2$-subunit relative to the $\alpha_1$-subunit
at the switch region\cite{Safo2005Hb}. This can be shown by
superimposing the C$_\alpha$ atoms of the $\alpha_1 \beta_1$ dimers and
computing the RMSD for the $\alpha_2 \beta_2$ dimer with the T$_0$
C$_\alpha$ atoms structure as a reference (Figure \ref{fig:struchb},
top panel) for all Hb forms extracted from crystal structures and for
the T$_0$ Hb structure simulated in different water box sizes. The
same applies to superimposing the C$_\alpha$ atoms of $\alpha_2
\beta_1$ dimers and computing the RMSD of the nonsuperimposed regions
($\alpha_1 \beta_2$ dimer, and also on the $\alpha$ carbons) with the
T$_0$ structure as a reference point (Figure \ref{fig:struchb}, left
bottom panel). And as a measure of quaternary variation, the complete
$\alpha_1 \beta_1$ $\alpha_2 \beta_2$ tetramer was superimposed on the
C$_\alpha$ atoms (Figure S2) 3rd panel from top). It is
found that these different RMSD results follow the same trends when
comparing different R forms and box sizes.  R$_3$ shows the most shift
and is closest to T$_0$, followed by RR$_2$ and lastly by the R$_2$
and R$_3$ structures.\\

\noindent	
Figure \ref{fig:struchb} also shows that the large quaternary
structural difference between the T and R forms is accompanied by
significant changes in the $\alpha_1$-$\alpha_2$ and
$\beta_1$-$\beta_2$ iron-iron distances; they are reduced in the
R-states, most notably for the R$_3$ structure (right panels). This
movement of the subunits has a large effect on the interdimer
interface (as observed in the interaction distances reported in Figure
S3) and thus on the central water cavity relative to
the T$_0$ structure. There is also the change in the
C$_\alpha$-C$_\alpha$ distance between His146$\beta_1$ and
His146$\beta_2$ and the C$_\alpha$-C$_\alpha$ distance between
His143$\beta_1$ and His143$\beta_2$, as well as the change in the
$\theta$ angle between the two planes containing
His146$\beta_1$-Fe$\beta_1$-Fe$\alpha_1$ and
His146$\beta_2$-Fe$\beta_2$-Fe$\alpha_2$. The angle between the
$\alpha_1 \beta_1$ and $\alpha_2 \beta_2$ subunits is smaller in all
R-forms compared to the T structure (Figure S2, bottom
panel).  These values explain the shorter distances between
His146$\beta_1$ and His146$\beta_2$, and between His143$\beta_1$ and
His143$\beta_2$ reported in Figure S2 (first two
panels from top). Hence, the $\beta$-cleft entrance to the central
water cavity is narrowed (compared to the T$_0$ structure with the
largest central cavity) and this leads to less water entering the
central cavity. The decrease in the number of water molecules in the
central cavity was noted in our previous paper \cite{MM.hb:2018} where
water molecules present in the central cylinder for the different box
sizes were counted (see Figures 5-figure supplement 3 and 4 in
Ref.\cite{MM.hb:2018}).\\

\noindent
Local structural changes around His146 resulting from differences in
the position of the $\beta_2$-subunit relative to the
$\alpha_1$-subunit are also observed (Figure S3). In all R
forms compared to the T structure, the water-mediated contact
(His146$\beta$)COO--OC(Pro37$\alpha$) and the salt bridges between
(His146$\beta$)COO--NZ(Lys40$\alpha$) and
(His146$\beta$)NE2--COO(Asp94$\beta$) are absent. Further, the
salt bridge missing in the T$_0$ form between (His146$\beta$)COO and
NE(His2$\beta$) is observed only in the R$_4$ form.\\

\noindent
Specific H-bonds at the $\alpha_1\beta_2$ dimer interface involved in
the shearing motion were also analyzed (Figure
S4). First, the hydrogen bond between Thr38$\alpha_1$
and His97$\beta_2$, present in the R$_3$ structure but missing in the
RR$_2$ and R$_2$ intermediate structures, was sampled in our
simulations. Second, the hydrogen bond between Tyr42$\alpha_1$ and
Asp99$\beta_2$ present only in the T$_0$ structure and absent in all
R-forms was observed for the stable T$_0$ state simulation (150 \AA\/
box) and was absent in all boxes with transitions. Finally, the
hydrogen bond between Arg92$\alpha_1$ and Gln39$\beta_2$ or
Glu43$\beta_2$ present in RR$_2$ and missing in all other states was
observed.\\

\begin{figure}[h!]
  \centering
  \includegraphics[scale=0.4]{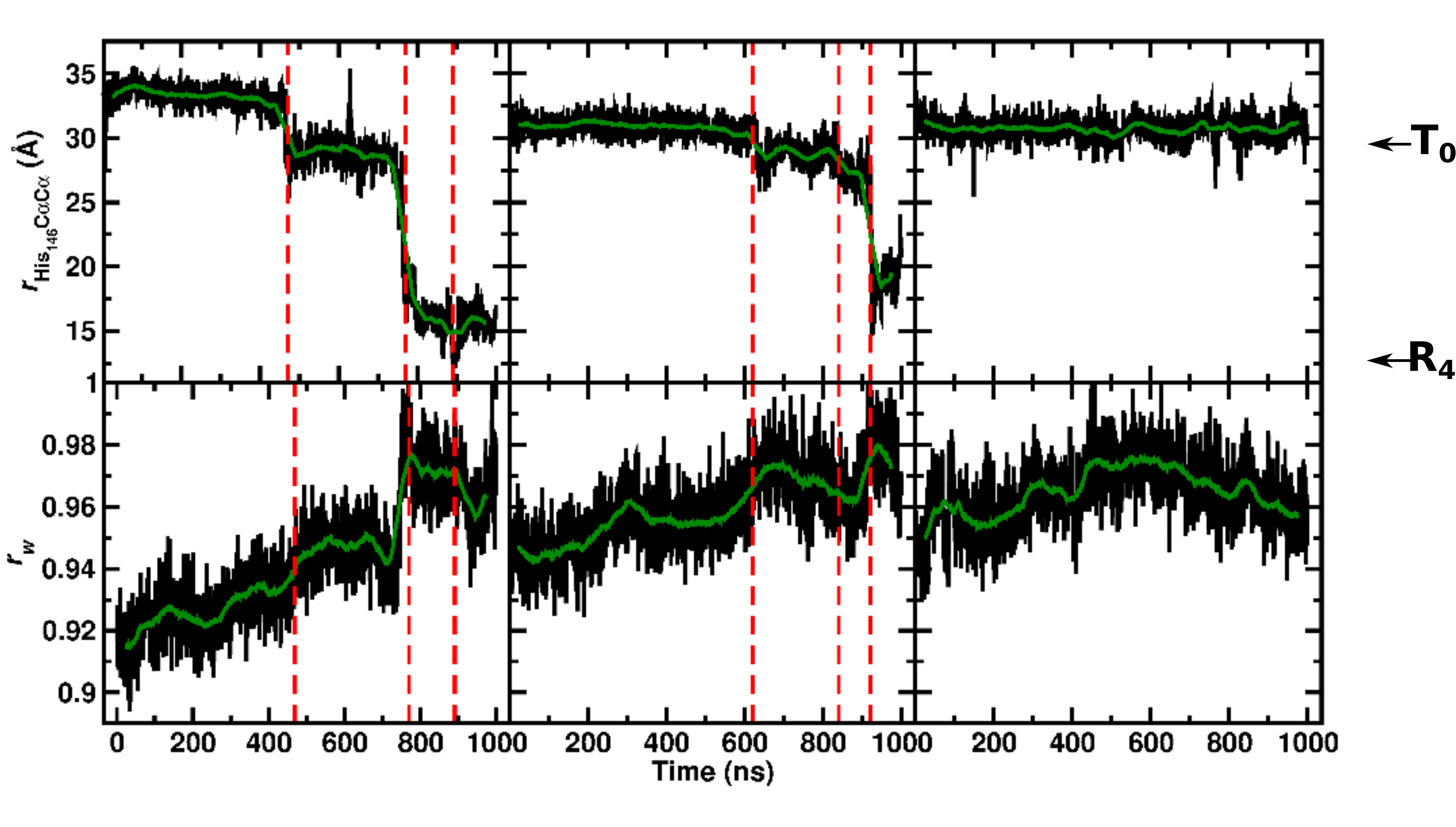}
\caption{Top: The C$_{\alpha}$ His$_{146} \beta_1$--His$_{146}
  \beta_2$ separation as a function of time for the 90, 120 and 150
  \AA\/ box from left to right. Raw data in black and running averages
  over 50 ns in green. The arrows on the right-hand side indicate the
  His$_{146}$$\beta_1$--His$_{146}$$\beta_2$ separation ($r_{\rm
    His146, C\alpha C\alpha}$) in the 2DN2 (T$_0$) and 2DN3 (R$_4$)
  crystal structures.\cite{tame:2006} Bottom: The hydration ("H$_2$O
  ratio") of the entire protein, expressed in terms of $r_w = N_{\rm
    wat} / N_{\rm max}$. The vertical red dashed lines indicate the
  transition times for the three steps observed in the 90 \AA\/ and
  120 \AA\/ boxes (see Figure 1B of reference \cite{MM.hb:2018}.}
\label{fig:step}
\end{figure}

\noindent
The conformational differences between the T and R states affect the
hydration environment in a manner that can be related to $\delta
\lambda_\mathrm{phob}^{(r)}$.  Based on the results of previous
simulations\cite{MM.hb:2018}, the set of residues for which $\delta
\lambda_\textrm{phob}$ changes most across the transitions was
selected. Figure \ref{fig:step} (top panel) reports the C$_{\alpha}$
His146$\beta_1$--His146$\beta_2$ separation, which serves as an
indicator of the T-to-R transition for the simulations in the 90, 120,
and 150 \AA\/ boxes. For the simulations in the two smaller boxes,
three red transitions are evident between the T$_0$-state (at early
times) and the R$_0$-state (at late times, see Figure 1 in
Ref.\cite{MM.hb:2018}), as indicated by the red dashed lines in
Figure~\ref{fig:step}. Structural changes are accompanied by changes
in the number of hydration waters. For the simulation in the largest
(150 \AA\/) box, for which no transition occurs, the C$_{\alpha}$
His146$\beta_1$--His146$\beta_2$ separation is constant and the
average hydration is larger than 0.95 (see bottom row in Figure
\ref{fig:step}). \\

\noindent
{\bf (a) Results for Hb 90 \AA\/ box: Local hydrophobicity (LH) for
  residues identified as the Perutz stereochemical model (see Table
  1).}  The LH analysis for the 1 $\mu$s simulation is carried out
with a time resolution of 0.5 ns. A cut-off of 6 \AA\/ from the
protein is chosen to distinguish between interfacial and bulk
water. The structural transitions for the 90 \AA\/ box occur at
$t=470$ ns, $t=770$, and $t=891$ ns, as indicated by the distance
$r_{\rm His146}$ between the C$_{\alpha}$ atoms of the two His146
residues in the $\beta_1$ and $\beta_2$ chains. The total number of
interfacial water is found to correlate with this distance (see bottom
row of Figure \ref{fig:step}). Whenever the distance between the two
His146 residues (see reference \cite{MM.hb:2018}) decreases abruptly
(as indicated by the red dashed lines), the relative number of water
molecules $r_w$ within the 6 \AA\/ cut-off increases. The value of $r_w
= N_{\rm wat} / N_{\rm max}$ was determined as the instantaneous
number $N_{\rm wat}$ of water molecules for a specific snapshot and
the maximum $N_{\rm max}$ encountered along the entire trajectory.\\

\begin{table}[h!]
\caption{The residues of Hb for which the local hydrophobicity $\delta
  \lambda_\textrm{phob}$ is analyzed from Perutz' stereochemical
  model\cite{perutz:1970}. For each residue its involvement in
  specific contacts is reported.}
\begin{tabular}{l|l}
\hline
Residue &  Role in the protein  \\
\hline 
Arg141$\alpha_1$     & $\alpha$ C-terminal salt bridge\\
Val1$\alpha_2$       & $\alpha$ C-terminal salt bridge\\
Asp126$\alpha_2$     & $\alpha$ C-terminal salt bridge\\
Lys127$\alpha_2$     & $\alpha$ C-terminal salt bridge\\
\hline
Tyr140$\alpha_1$     & $\alpha$  proximity to the C-terminal residue \\
\hline
His146$\beta_1$      & $\beta$ C-terminal salt bridge\\
Lys40$\alpha_2$      & $\beta$ C-terminal salt bridge\\
Asp94$\beta_1$      & $\beta$ C-terminal salt bridge\\
\hline
Tyr145$\beta_1$    & $\beta$  salt bridge involved in His146$\beta_1$ motion\\
Val98$\beta_1$      & $\beta$  salt bridge involved in His146$\beta_1$ motion\\

\hline
Thr38$\alpha_1$      & $\alpha_1 - \beta_2$ shearing \\
Thr41$\alpha_1$       & $\alpha_1 - \beta_2$ shearing\\
Tyr42$\alpha_1$        & $\alpha_1 - \beta_2$ shearing\\
Asp94$\alpha_1$        & $\alpha_1 - \beta_2$ shearing\\
Cys93$\beta_2$         & $\alpha_1 - \beta_2$ shearing\\
Val98$\beta_2$       & $\alpha_1 - \beta_2$ shearing\\
Asn102$\beta_2$      & $\alpha_1 - \beta_2$ shearing\\
Asp99$\beta_2$      & $\alpha_1 - \beta_2$ shearing\\
\hline
\label{tab:tab1}
\end{tabular}
\end{table}

\begin{figure}[h!]
\centering
\includegraphics[scale=0.6]{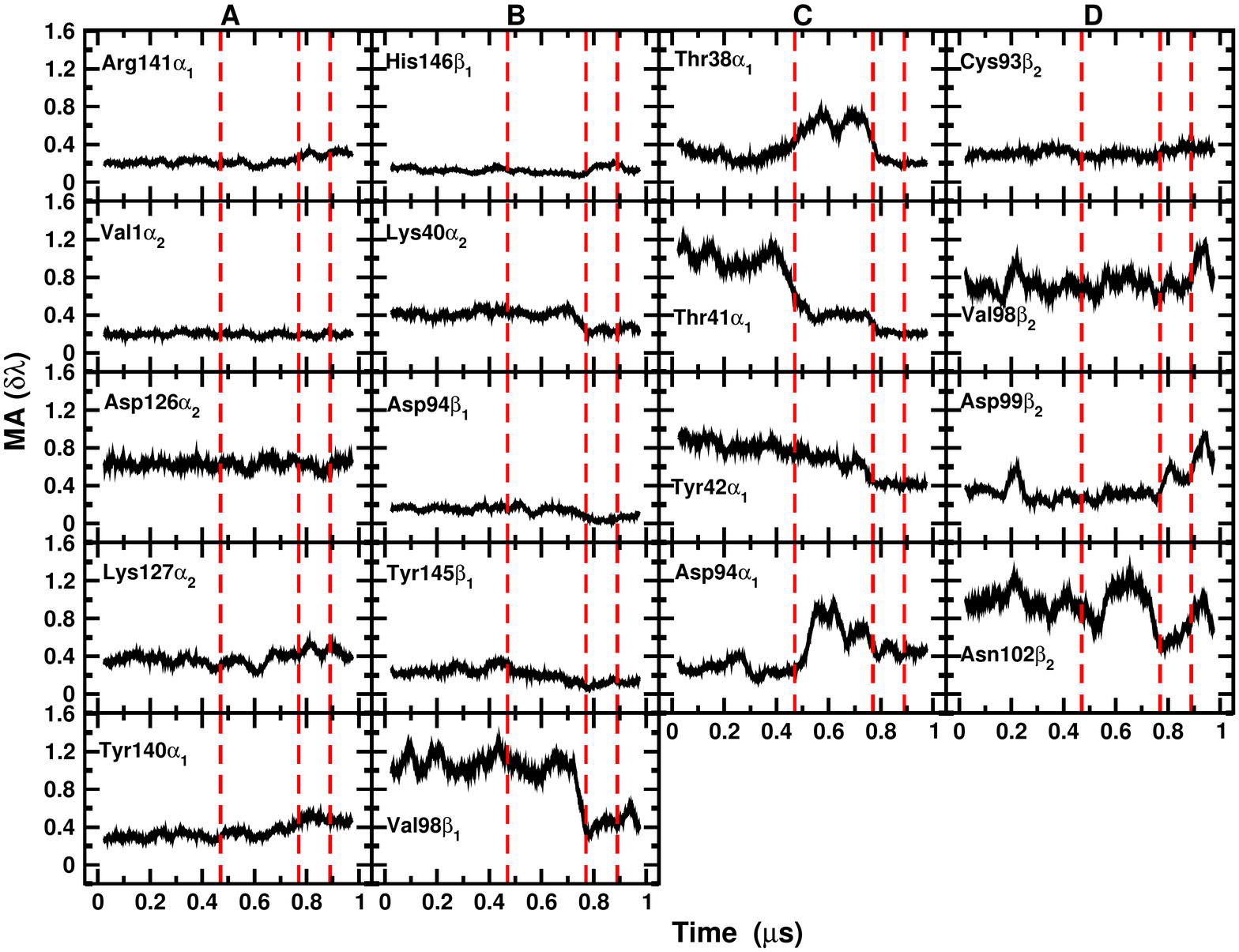}
\caption{Local hydrophobicity for the simulation in the 90 \AA\/
  box. Moving average (MA) over 50 ns of $\delta \lambda_\textrm{phob}$ as
  a function of time for residues involved in the C-terminal salt
  bridges\cite{perutz:1970} (column A), additional salt bridges
  (column B) and the $\alpha_1 \beta_2$ 
(columns C and D); see Table \ref{tab:tab1}.}
\label{fig:fig90box}
\end{figure}

\noindent
To obtain more detailed information, $\delta \lambda_\textrm{phob}(t)$
was analyzed for the residues listed in Table \ref{tab:tab1}, see
Figure \ref{fig:fig90box}. For certain residues, structural
transitions (at $t = 470$ ns, $t = 770$ ns, and $t = 891$ ns) are
accompanied by abrupt rather than by gradual changes in local
hydrophobicity of individual residues. Examples include
Val98$\beta_1$, Thr41$\alpha_1$, or Asp94$\alpha_1$. By contrast,
Tyr42$\alpha_1$ shows a gradual decrease in $\delta
\lambda_\mathrm{phob}$ over most of the 1 $\mu$s simulation. There are
also changes in LH away from overall structural transitions, e.g. for
Val98$\beta_2$, Asp99$\beta_2$, and Asn102$\beta_2$ at 200 ns, further
discussed below. Except for Val98$\beta_1$ all residues that show a
substantial decrease in their hydrophilic ($\delta
\lambda_\textrm{phob} \sim 0.5$) versus hydrophobic ($\delta
\lambda_\textrm{phob} \sim 0$) character [Thr41$\alpha_1$,
  Tyr42$\alpha_1$] or an increase [Thr38$\alpha_1$, Asp94$\alpha_1$,
  Asp99$\beta_2$] are at the $\alpha_1 / \beta_2$ interface. This
suggests that the decay for the 90 \AA\/ box is triggered primarily by
the hydration around residues that are involved in the $\alpha_1 /
\beta_2$ contacts.\\
 
\noindent
Based on the data in Figure~\ref{fig:fig90box}, the T$_0
\rightarrow$R$_0$ transition in the 90 \AA\/ box is accompanied by
significant changes in the hydration environment at certain locations
around the $\alpha_1 / \beta_2$ contact.  This observation is
consistent with Perutz' conclusion. We quote, ``[..]Where is the force
that changes the quaternary structure applied[..]The evidence is
overwhelmingly in favor of the contacts $\alpha_1
\beta_2$[..]''.\cite{perutz:1970} The significant changes in hydration
around the $\alpha_1 \beta_2$ contact suggest the possibility that the
T$_0 \rightarrow$R$_0$ transition is driven by solvent
thermodynamics.\\

\begin{figure}[h!]
\centering \includegraphics[scale=0.6]{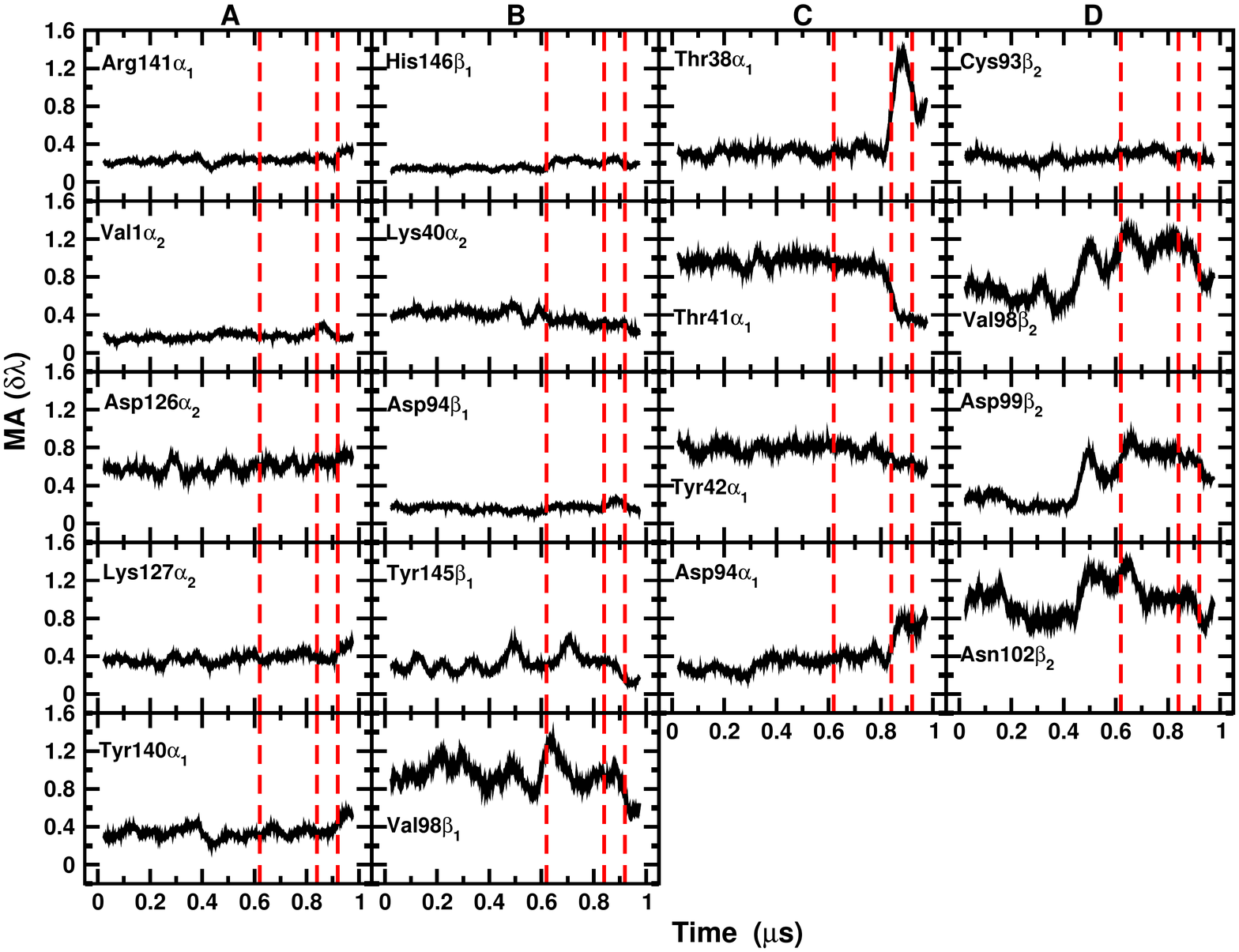}
\caption{Local hydrophobicity for the simulation in the 120 \AA\/
  box. Moving average [MA ($\delta \lambda$)] over 50 ns of $\delta
  \lambda_\textrm{phob}$ as a function of time for residues involved
  in the C-terminal salt bridges\cite{perutz:1970} (column A),
  additional salt bridges (column B) and the $\alpha_1 \beta_2$
  shearing motion (columns C and D); see Table \ref{tab:tab1}.
The presence of the R$_2$ structure at 920 ns is indicated in all the panels
by a vertical violet dashed line.}
\label{fig:fig120box}
\end{figure}

\noindent
{\bf (b) Results for Hb 120 \AA\/ box:} For the simulation in the 120
\AA\/ box most of the residues involved in the salt bridges, like
those in the 90 \AA\/ box, show only minor variations in $\Delta
\langle \delta \lambda_\textrm{phob} \rangle$ except that of
Tyr145$\beta_1$ and Val98$\beta_1$ which have the largest variations
along the trajectory (see Figure \ref{fig:fig120box}B). It is found
that the LHs of all other residues in Figures \ref{fig:fig90box}A, B
and \ref{fig:fig120box}A, B behave similarly in the simulations of the
90 \AA\/ and 120 \AA\/ boxes. For Val98$\beta_1$, instead of decaying
to $\Delta \langle \delta \lambda_\textrm{phob} \rangle \approx 0$ as
in the simulation of the 90 \AA\/ box, the value of $\Delta \langle
\delta \lambda_\textrm{phob} \rangle$ in the 120 \AA\/ box remains at
or above 0.5 throughout the entire simulation. Hence, the T$_0$
$\rightarrow$ R$_0$ transition is again mainly associated with motion
at the $\alpha_1 / \beta_2$ interface (see Figure
\ref{fig:fig120box}C,D). An example of a transition that follows the
mechanism described by Perutz\cite{perutz:1970} is the transition at
620 ns at the $\alpha_2/\beta_1$ interface, where the cleavage of the
Tyr42$\alpha_2$ and Asp99$\beta_1$ salt bridge is clearly visible (see
Figure S4 middle panel and Figure
S5).\\

\noindent
Several of the residues at the $\alpha_1 / \beta_2$ interface show
pronounced changes in local hydrophobicity that coincide with
structural transitions (Figures S3 and
S4). However, a few residues in Figure
\ref{fig:fig120box}D also show LH changes that are not necessarily
linked directly to a tertiary structural change (``step''); they are
Val98$\beta_2$, Asp99$\beta_2$, and Asn102$\beta_2$ at around 500 ns,
further discussed below. For the transitions at 620 ns and 840 ns
there is again a clear change in LH for Thr38$\alpha_1$,
Thr41$\alpha_1$, and Asp94$\alpha_1$, the most pronounced of them
involving Thr38$\alpha_1$. These observations also indicate that the
nature of the transition at 470 ns (step1) in the 90 \AA\/ and at 620
ns (step1) in the 120 \AA\/ box is different and may be explained by
the transition to different intermediate R-forms (see next
paragraph).\\

\noindent
For the decaying structures in the 90 \AA\/ and 120 \AA\/ boxes the
following is observed. In the 90 \AA\/ box, a transition from T$_0$ to
R$_3$ starts at 470 ns (step1), where the His146--His146 separation
drops from 31 \AA\/ to 25 \AA\/ bringing the Hb structure closer to
R$_3$ (22 \AA\/, Figure S2, first panel). This T$_0$
to R$_3$ transition is completed at 780 ns (step2, Figure
S2, the presence of the R$_3$ structure at 780 ns is
marked in all the panels by a cyan dashed line). At $\sim 880$ ns
(step3) a next transition from R$_3$ to R$_2$ occurs (the R$_2$
structure is marked by a violet dashed line in Figure
S2). For the rest of the simulation until 1 $\mu$s,
the RR$_2$ and R$_4$ states are sampled (Figure S2,
RR$_2$ and R$_4$ states are indicated by a green dashed line and a
blue arrow, respectively). Conversely, in the 120 \AA\/ box, at step1
at 620 ns a T$_0$ to R$_2$ transition starts by decaying to an unknown
intermediate. It is continued at 840 ns (step2) to bring the structure
closer to R$_2$. The transition to R$_2$ is completed by 920 ns
(step3, Figure S2), the presence of the R$_2$
structure at 920 ns is indicated in all the panels by a vertical violet
dashed line).\\

\begin{figure}[h!]
\centering
\includegraphics[scale=0.6]{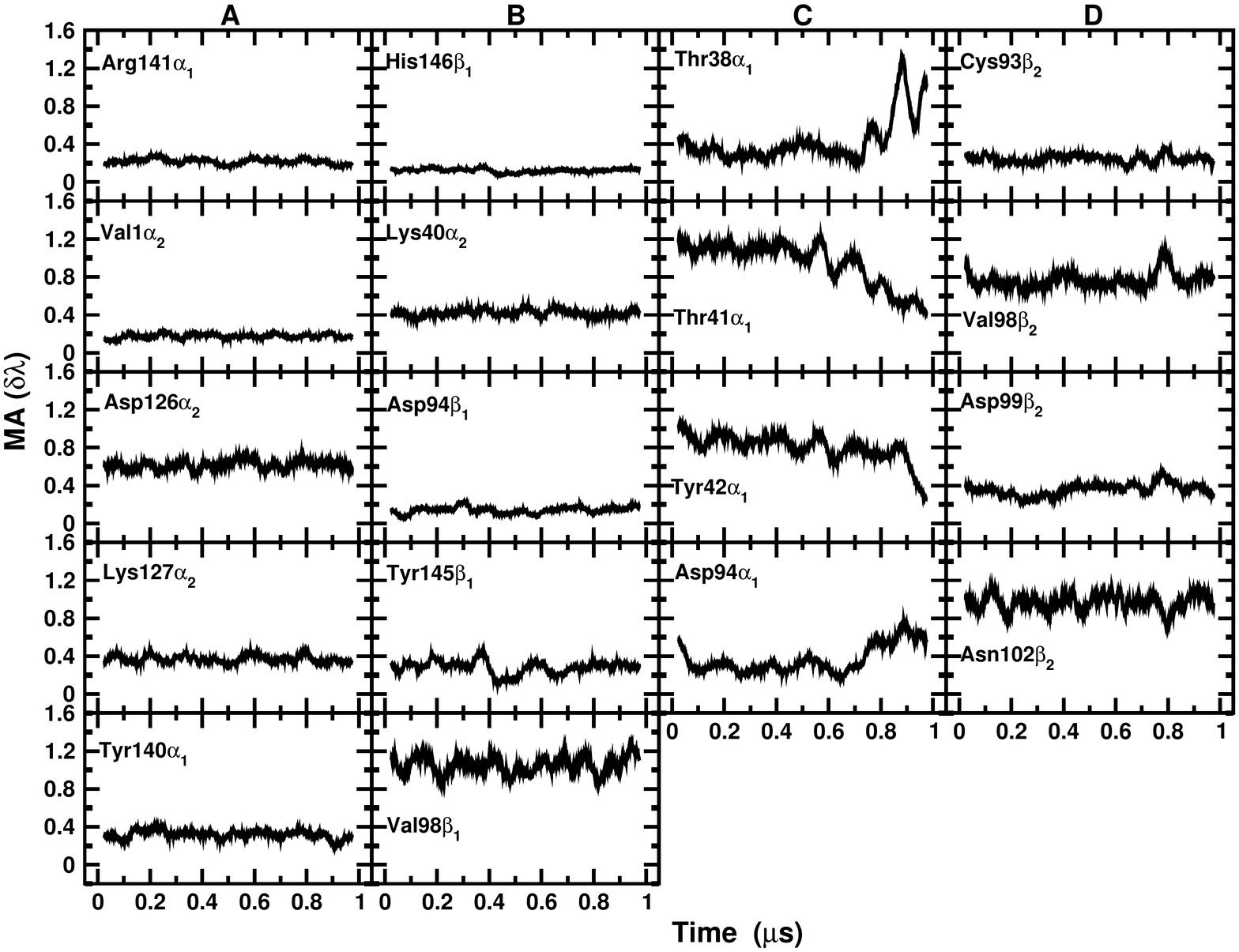}
\caption{Local hydrophobicity for the simulation of the 150 \AA\/ box.
  It shows moving averages over 50 ns of $\delta
  \lambda_\textrm{phob}$ as a function of time for residues involved
  in the C-terminal salt bridges\cite{perutz:1970} (column A),
  additional salt bridges (column B) and the $\alpha_1 \beta_2$
  shearing motion (columns C and D), see Table \ref{tab:tab1}.}
\label{fig:fig150box}
\end{figure}

\noindent
{\bf (c) Results for Hb 150 \AA\/ box:} For the 150 \AA\/ box the
previous MD simulations did not find a structural
transition.\cite{MM.hb:2018} The values of $\delta
\lambda_\textrm{phob}$ for all residues involved in the salt bridges
(Figure \ref{fig:fig150box}, columns A and B) do not deviate
significantly from their average value. The amplitude of the
fluctuations are typically smaller than for the simulations in the 90
\AA\/ and 120 \AA\/ boxes. For the residues at the $\alpha_1 \beta_2$
interface there are variations for Thr38$\alpha_1$, Thr41$\alpha_1$,
Tyr42$\alpha_1$, and Asp94$\alpha_1$ (see Figure
\ref{fig:fig150box}).\\

\noindent
The clearest difference between the simulation in the 150 \AA\/ and
the two smaller boxes is the behaviour for residues Val98$\beta_2$,
Asp99$\beta_2$, and Asn102$\beta_2$. As an example, the water
occupation around Val98$\beta_2$ is analyzed by computing the radial
distribution function $g(r)$ between the C$_{\alpha}$ of the residue
and the surrounding hydration water for different parts of the
trajectory. The radial distribution functions $g(r)$ in Figure
S6 show that they are close in shape to one another
but differ in magnitude for the early phase of the trajectory in the
120 \AA\/ and 150 \AA\/ box.  They change in shape after the
transition at 840 ns in the smaller of the two boxes. A pronounced
signature in LH is also found in the 120 \AA\/ and 150 \AA\/ boxes for
Thr38$\alpha_1$ between 800 and 900 ns. The signatures in LH for the
150 \AA\/ box can be related to formation of a
Thr38$\alpha_1$--Asp99$\beta_2$ salt bridge (Figure
S7). Breaking and reforming of salt bridges
involving Val98$\beta_2$, Asp99$\beta_2$, and Asn102$\beta_2$ is also
responsible for the sharp increase in LH around these three residues
in the 90 \AA\/ box around 200 ns, see Figures \ref{fig:fig90box}D and
S8. It is notable that the LH around the three
residues already starts to change before the salt bridge actually
breaks. \\

\noindent
Changes of the LH of each residue can be due to a) internal motion of
a residue or b) the influence of neighbouring residues. Both of these
are potentially followed by water displacement or influx which change
$\delta \lambda_{\rm phob}$. For the transition at $t=840$ ns in the
120 \AA\/ box, changes in the local water occupation around
Val98$\beta_2$ obtained by analysis of radial distribution functions
(see Figure S6) do not necessarily lead to changes
in LH. The $g(r)$ for the time intervals 0 to 620 ns and 620 ns to 840
ns are very similar (red and green lines in Figure
S6) while the LH changes from 0.8 at early times to
1.2 after $t \sim 500$ ns, see Figure \ref{fig:fig120box}. The water
influx is a consequence of the reconfiguration of the H-bonding
network including the Tyr42$\alpha_1$--Asp99$\beta_2$ salt bridge (see
Figure S9) and the rearrangement of the carboxy group
of the sidechain of Asp99$\beta_2$ due the rehydration of the side
chain. These effects are also mirrored by the Asp99$\beta_2$ carboxy
orientation (see dihedral time series reported in Figure
S11) which demonstrates that before the transition at
840 ns the side chain follows a two-state behaviour but after the
transition almost free rotation occurs (see also Figure
S4). This change is accompanied by increased hydration
of the side chain (bottom panel of Figure S11).\\

\noindent
Comparing Figures \ref{fig:fig90box} to \ref{fig:fig150box} it is
noted that even when Hb is still in its T$_0$ state (i.e. before 470
ns, the first transition in the 90 \AA\/ box), differences in LH,
mainly at the $\alpha_1 / \beta_2$ interface are observed. Examples
include residues Thr41$\alpha_1$ and Tyr42$\alpha_1$ for which LH
oscillates or decreases in the 90 \AA\/ box but remains constant in
the two larger boxes before 470 ns. The finding that destabilization
of the $\alpha_1 / \beta_2$ interface is at the origin of the T$_0$ to
R$_0$ transition is consistent with the Perutz stereochemical
model. Conversely, the LH around the C-terminal salt bridge residues
is very similar for the simulations in the three different box sizes,
except for Val98$\beta_1$ and Tyr145$\beta_2$.\\

\noindent
{\bf (d) Spatio-temporal analysis based on two-dimensional correlation
  maps:} To better understand the coupling of local hydration dynamics
and the structural transitions, two-dimensional correlation maps were
generated which are referred to as local hydrophobicity cross
correlation maps (LH-CCMs). Similar to dynamic cross correlation maps
(DCCMs)\cite{karplus:1991,karplus:1996} for residues $i$ and $j$ the
quantity
\begin{equation}
  C_{ij} = \frac{\langle \Delta \delta\lambda_\textrm{phob}^{(i)}
    \Delta \delta\lambda_\textrm{phob}^{(j)} \rangle}
  {\sqrt{\langle (\Delta \delta\lambda_\textrm{phob}^{(i)})^2 \rangle
\langle (\Delta \delta\lambda_\textrm{phob}^{(j)})^2 \rangle }
  }
\label{eq:dccm}
\end{equation}
was determined for each interval for which Hb was in a particular
conformational state as shown in Figure \ref{fig:struchb}.\\

\begin{figure}
\centering
\includegraphics[scale=0.50]{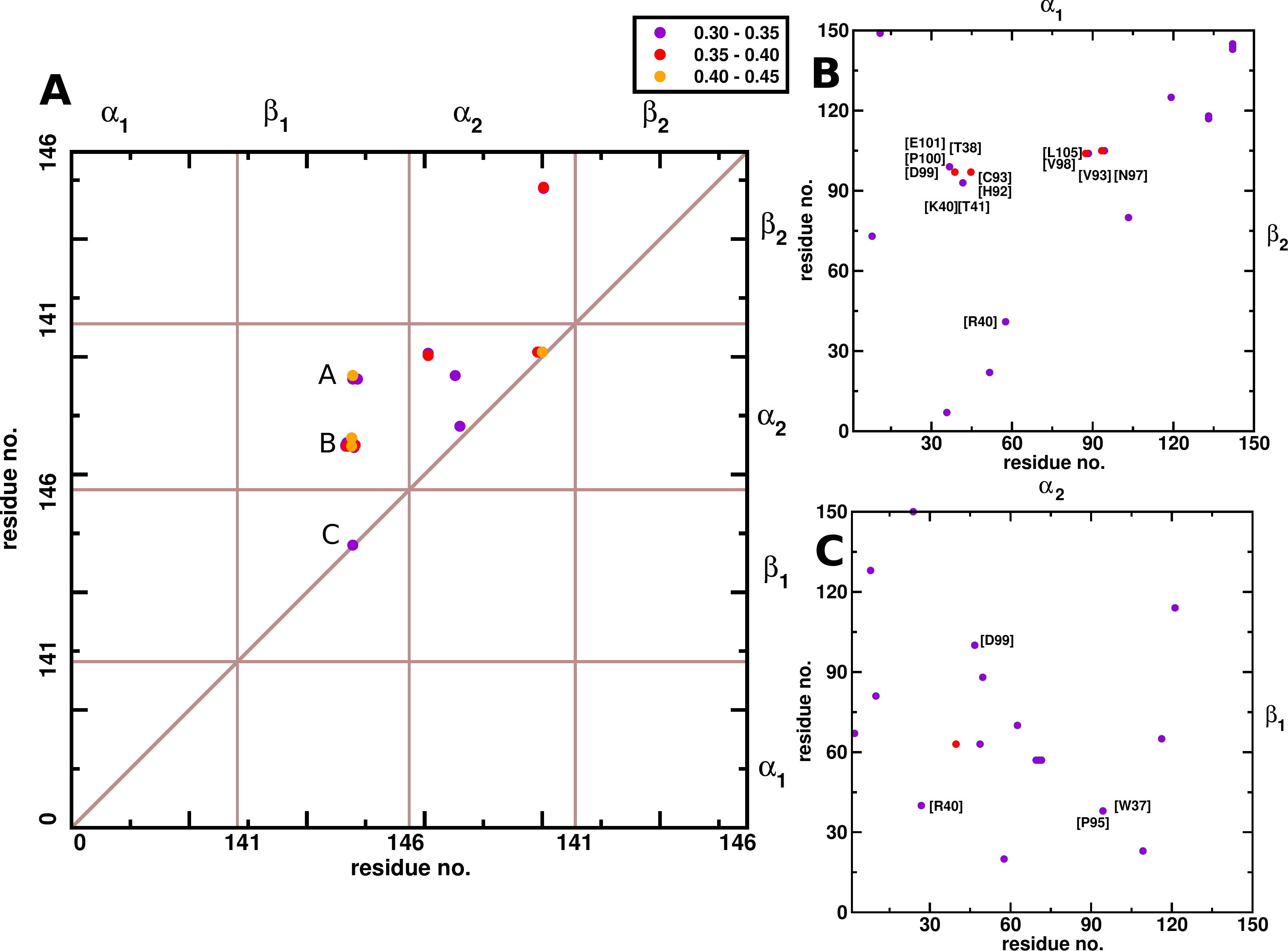} 
\caption{Difference of Local Hydrophobicity $\Delta C_{ij}$ Cross
  Correlation Map (LH-CCM) for the 120 \AA\/ box. Panel A: difference
  between [0-620] and [620-840], i.e. "transition at 620" for entire
  protein. Panel B: difference between [620-840] and [840-920],
  i.e. "transition at 840" for the $\alpha1 / \beta2$ interface. Panel
  C: difference between [620-840] and [840-920], i.e. "transition at
  840" for the $\alpha2 / \beta1$ interface. The cross correlations
  for the individual states are given in Figure
  S12. Only values $\Delta C_{ij} \geq 0.30$ are
  reported. Color code: $0.30 \leq \Delta C_{ij} < 0.35$ (purple),
  $0.35 \leq \Delta C_{ij} < 0.40$ (red), and $0.40 \leq \Delta C_{ij}
  < 0.45$ (orange).}
\label{fig:lh120diff} 
\end{figure}

\noindent
Figure \ref{fig:lh120diff}A reports the difference between the local
hydrophobicity cross correlation maps between time intervals 0 to 620
ns and 620 to 840 ns for the 120 \AA\/ box for values of $\Delta
C_{ij} > 0.3$. This map indicates that correlations in LH and their
difference can depend on both the sequence and spatial proximity of
two residues. The correlation in LH up to 620 ns (i.e. before the
first transition, see Figure S10) is large ($C_{ij} >
0.30$) for residues that play an active role in interface transitions
between the two subunits (Figure \ref{fig:lh120diff} above the
diagonal) and for regions that are spatially close (on the
diagonal). An example for sequence proximity is the
Val98$\beta_1$-Asp$99\beta_1$ region (feature C in Figure
\ref{fig:lh120diff}A and Figure S9). Changes in LH are
a direct consequence of the Asp99$\beta_1$-Tyr42$\alpha_2$ salt bridge
cleavage during the transition at 620 ns (see Figure
S5) that leads to a change in the orientation of the
peptide bond between Val98$\beta_1$-Asp$99\beta_1$ (Figure
S9) and corresponding decrease in the LH of both,
Asp99$\beta_1$ and Tyr42$\alpha_2$.  Examples for spatial proximity of
residues are the two clusters (labelled A and B in Figure
\ref{fig:lh120diff}A) that are at the $\alpha_2 \beta_1$ shearing
interface. This change in hydrophobicity in one of the two important
stabilizing regions of the protein (the other being the $\alpha_1 /
\beta_2$ interface), indicates its possible involvement in the
destabilization of the T$_0$ structure.\\

\noindent
A more detailed view of the LH cross correlations for the $\alpha_1
\beta_2$ interface for the 120 \AA\/ box is provided in Figure
S10 (for LH-CCMs in the 90 \AA\/ and 150 \AA\/ boxes
see Figures S13 and S14).  Figure
S10 shows the LH cross correlations for residues
involved in the $\alpha_1 / \beta_2$ shearing motion up to 620 ns. The
clusters (A to E) in Figure S10 involve correlated
changes in LH at the $\alpha_1$/$\beta_2$ interface, whereas for
cluster F no direct contact is present. In cluster A the correlation
is caused by the Thr41$\alpha_1$--Arg40$\beta_2$ salt bridge present
before the transition at 620 ns. Cluster B is dominated by the
water-mediated Asp94$\alpha_1$--Arg40$\beta_2$ salt bridge before the
transition at 620 ns; it is a weak interaction due to the large
distance ($\sim 6$ \AA\/) between the proton and the anion. After the
transition this salt bridge becomes the dominant interaction in which
Arg40$\beta_2$ is involved. The C cluster is dominated by the
$\pi$-stacking interaction between Tyr140$\alpha_1$ and
Trp37$\beta_2$. A weak salt bridge of the Thr38$\alpha_1$ and
Asp99$\beta_2$ sidechains with the Thr41$\alpha_1$ sidechain and
Asp99$\beta_2$ NH peptide bond are responsible for the D cluster. The
NH peptide bond of Asn97$\alpha_1$ and the Asp99$\beta_2$ side chain
lead to cluster E. Overall, this figure provides a dynamic view of the
stereochemical model proposed by Perutz.\cite{perutz:1970} This is
illustrated, for example, by the fact that all clusters (A to F) are
extended, rather than the usual point-to-point contacts (i.e., the
salt bridges) alone.\\

\noindent
Next, the transition in the 120 \AA\/ box at 840 ns is discussed from
the perspective of the LH-CCMs (see Figure \ref{fig:lh120diff} panels
B and C). They show the difference between the LH-CCMs for the time
intervals [620-840] ns and [840-920] ns, respectively. During the
process two salt bridges are broken (Thr41$\alpha_1$--Asp99$\beta_2$
and Thr41$\alpha_1$-Arg40$\beta_2$, which is water mediated) and two new
salt bridges are formed (Thr38$\alpha_1$--Asp99$\beta_2$ and
Asp94$\alpha_1$--Arg40$\beta_2$) and Asn97$\alpha_1$ -
Asp99$\beta_2$ continues to show a bimodal behaviour, see Figure
S15. It is found that the reformation of these
salt bridges between residues involved in the ``Perutz
mechanism'' (Thr38$\alpha_1$, Thr41$\alpha_1$, Asp94$\alpha_1$ and
Asp99$\beta_2$) is also reflected in the difference cross correlation
maps (Figure \ref{fig:lh120diff}B and C).  They confirm that most of
the changes for this transition occur at the $\alpha_1 / \beta_2$
interface. Also, these two panels show that changes in the LH-CCMs are
not necessarily symmetric for the $\alpha_1 / \beta_2$ and $\alpha_2 /
\beta_1$ interfaces. Such a ``dynamical asymmetry'' 
(i.e., it is found in the molecular dynamics simulations) has 
also been observed for insulin dimer\cite{MM.insulin:2019} for which
the X-ray structure has C$_2$ symmetry\cite{cutfield:1979} or is very
close to symmetric with only small local deviations from
it.\cite{baker:1988}\\

\begin{figure}[h!]
\centering
\includegraphics[scale=0.75]{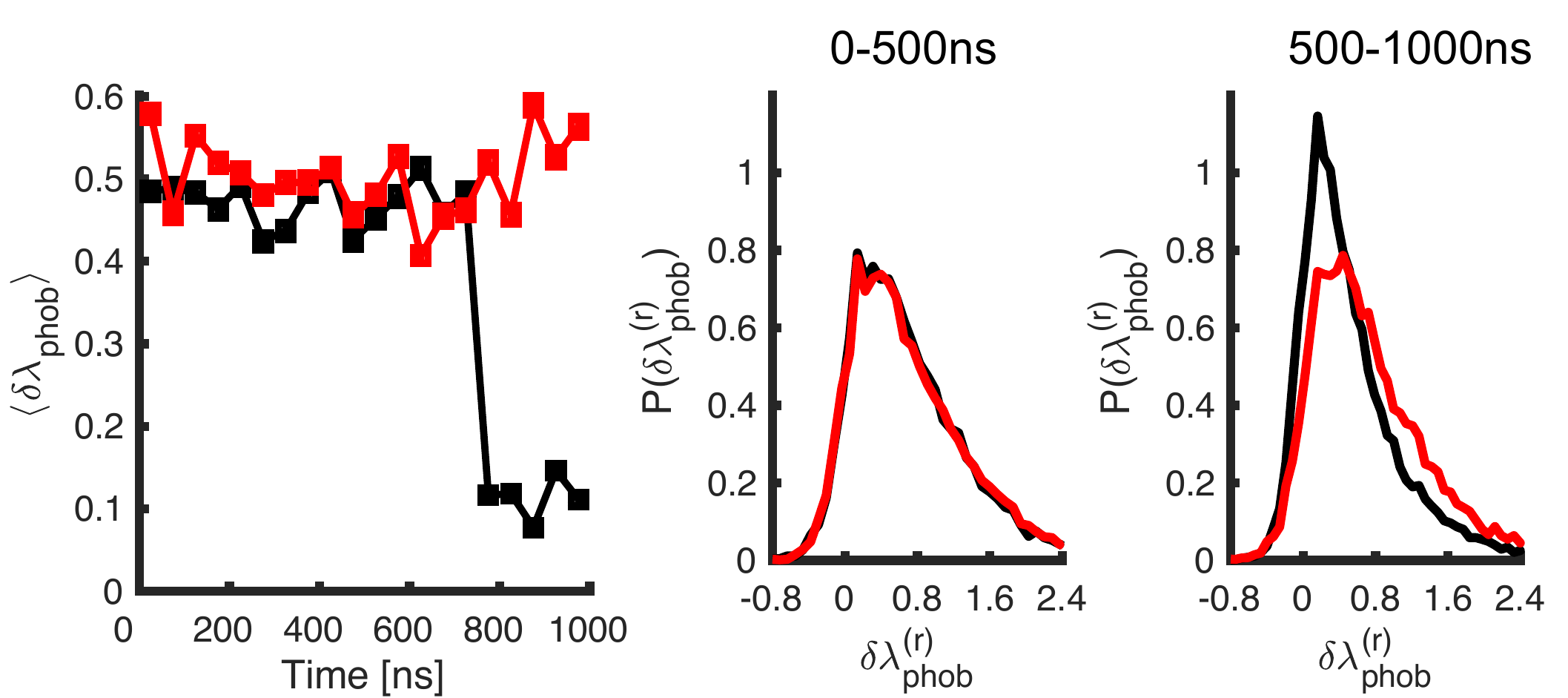} 
\caption{Comparison of the interfacial water structure for residues
  that are part of the $\alpha_1/\beta_2$ and $\alpha_2/\beta_1$
  interfaces. The black and red lines correspond to simulations
  carried out in $90 \mathrm{\AA}$ and $150 \mathrm{\AA}$ solvent
  boxes, respectively. The marker indicates the center of a time
  interval. Panel A: The expectation value of
  $\delta\lambda_\mathrm{phob}^{(r)}$ as a function of time, averaged
  over 50 ns time intervals.  Panel B: The distribution of $\delta
  \lambda_\mathrm{phob}^{(r)}$ values during the first 500 ns of the
  trajectories.  Panel C: The distribution of
  $\delta\lambda_\mathrm{phob}^{(r)}$ values during the last 500 ns of
  the trajectories.}
\label{fig:shinv2}
\end{figure}

\noindent
As previous results have shown, the relative stability of the T$_0$
state depends on the size of the simulation cell.\cite{MM.hb:2018}
Analysis of hydration structure via $\delta
\lambda_\mathrm{phob}^{(r)}$ has the ability to reveal when and where
protein hydration properties differ between differently sized
simulation cells. To highlight this point, the statistics and dynamics
of $\delta \lambda_\mathrm{phob}^{(r)}$ for Hb in the T$_0$ state in
the 90 \AA\/ and 150 \AA\/ simulation boxes are compared in
Figure~\ref{fig:shinv2}. It summarizes the values of $\delta
\lambda_\mathrm{phob}^{(r)}$ over the residues that comprise the
$\alpha_1/\beta_2$ and $\alpha_2/\beta_1$ interfaces, as a function of
simulation time. Most notably, as the conformational transition from
the $T_0$ state to the $R_0$ state progresses, there is a distinct
shift towards values of $\delta\lambda_\mathrm{phob}^{(r)}$ near
zero. This indicates that the interfacial water structure shifts from
that observed at a hydrophilic surface towards that observed at a
hydrophobic surface (Figure~\ref{fig:shinv2}A). The shift in
interfacial water structure is also apparent from the probability
distribution, $P(\delta \lambda_\mathrm{phob}^{(r))})$, for the
different simulation box sizes. During the first 500ns of the
trajectories the probability distributions overlap
(Figure~\ref{fig:shinv2}B), showing that the interfacial water
structure does not differ significantly. However, during the last 500
ns of the trajectories the probability distribution for the 90\AA\/
simulation box has a significant shift towards lower values of $\delta
\lambda_\mathrm{phob}^{(r)}$, corresponding to a more hydrophobic
character of the interface (Figure~\ref{fig:shinv2}C).\\

\subsection{Hydration Dynamics around Melittin}
\begin{figure}
\centering
\includegraphics[scale=0.34]{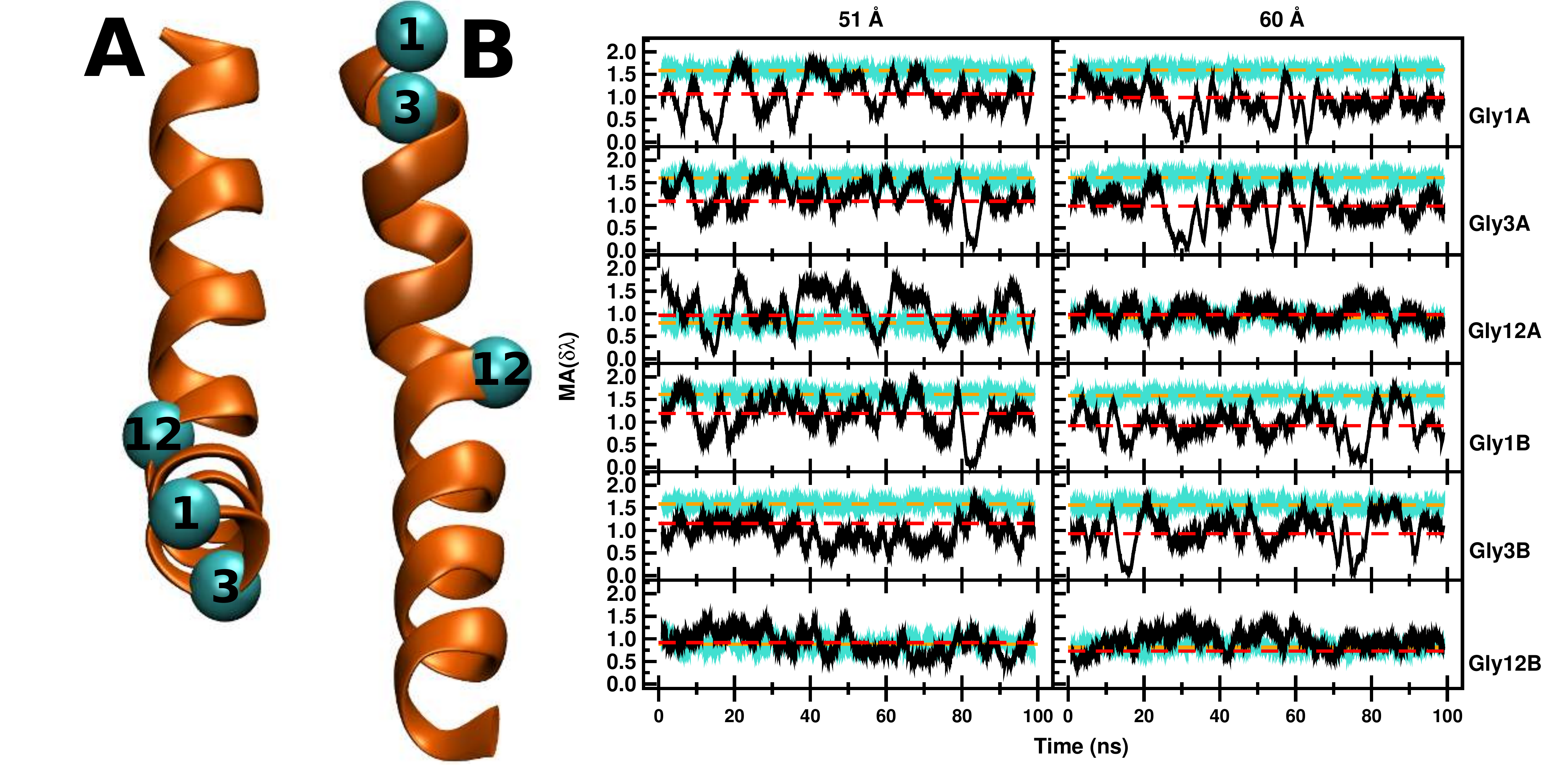} 
\caption{Left: A- and B-helices with the Gly1, Gly3, and Gly12
  residues represented as spheres. The A-helix is slightly bent at the
  C-terminal end for the rigid and flexible simulations. The RMSD of
  the two chains of the dimer is 1.6 \AA\/. Right: Time evolution of
  $\delta \lambda_\textrm{phob}$ for the 6 glycine residues in the
  melittin dimer. The results of the simulations for the rigid (cyan)
  and flexible (black) protein are reported. The orange and red line
  represent the average $\delta \lambda_\textrm{phob}$ for each
  residue during the simulation. Gly12 is less hydrophilic than the
  other Gly residues since it is located in the hydrophobic region of
  the protein.}
\label{fig:fig1}
\end{figure}

As a second example, the analysis of water hydration for Hb has been
extended to melittin.  It is a well studied prototype of a protein
complex that is stabilized through hydrophobic interactions. Melittin
is a small, 26-amino acid protein (for the sequence, see Figure
S16) found in honeybee venom that crystallizes as a
tetramer, consisting of two dimers (see Figure \ref{fig:fig1} left),
related by a two-fold symmetry
axis.\cite{terwilliger1982.1,terwilliger1982.2} Cheng and
Rossky\cite{rossky:1998} characterized the behaviour of the
hydrophobic surface of the melittin dimer and of the surrounding
surface residues by simulations in which the structure of the melittin
dimer was frozen. They found that in hydrophilic regions the water
molecules have a well defined orientation, while in the hydrophobic
regions, the waters are more mobile and explore different
configurations.\cite{rossky:1998} To further explore the hydration
dynamics, simulations with a frozen melittin dimer in different box
sizes are carried out and analyzed. In additional simulations, the
protein was also allowed to move freely.  These provide information
about the solvent-solute coupling.\\

\noindent
It is of interest to analyze whether $\delta
\lambda_{\rm phob}$ encapsulates corresponding information, and
whether simulations of water around a rigid melittin dimer, as carried
out in Ref.\cite{rossky:1998}, and around a flexible dimer lead to
qualitatively similar results. Since the results for Hb
depend on the box size, simulations are also carried out with
different box sizes.\\

\begin{figure}[h!]
\centering
\includegraphics[scale=0.55]{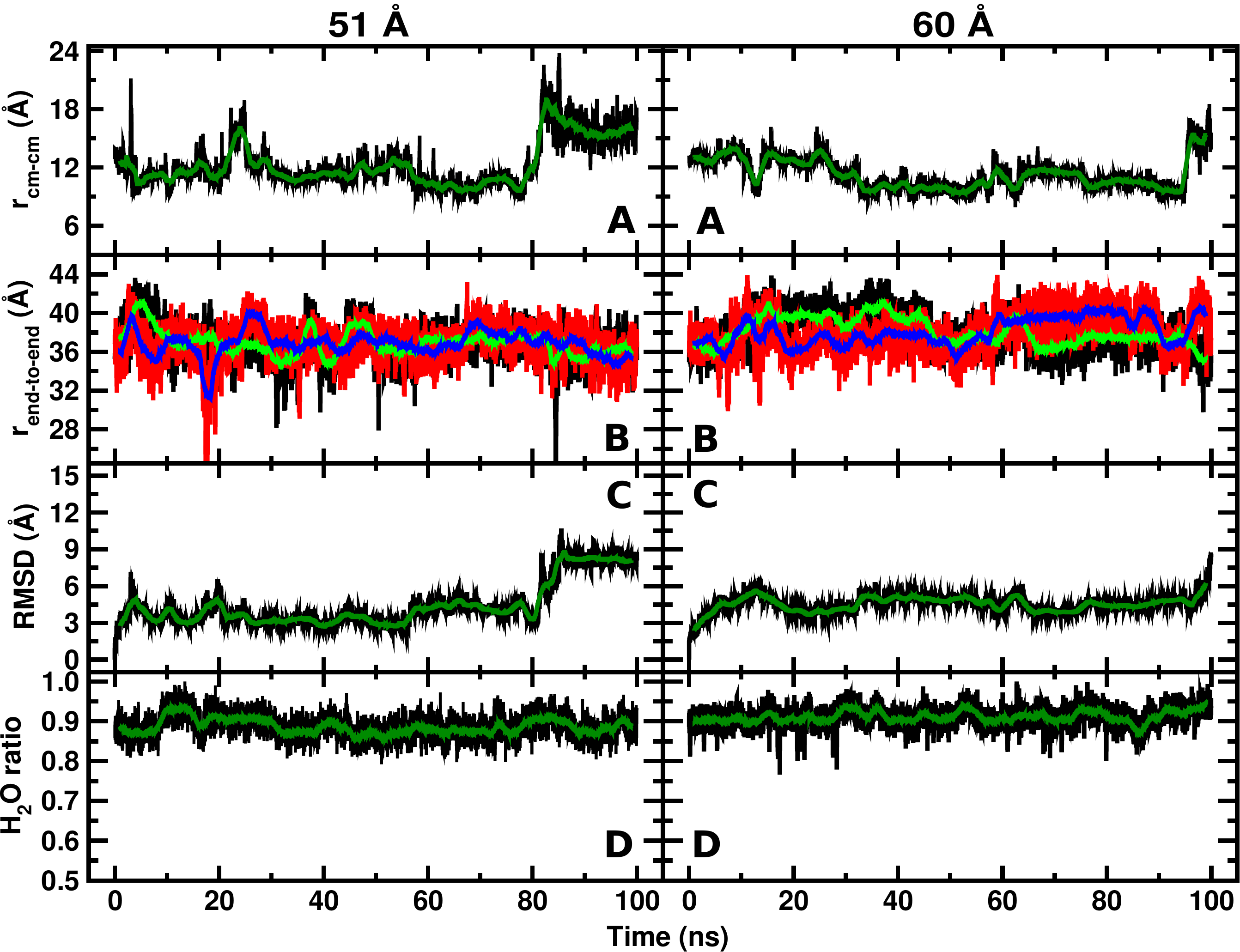} 
\caption{Flexible melittin in the 51 \AA\/ and 60 \AA\/ boxes. In
  Panels A the center-of-mass distance between the two monomers
  ($r_{\textrm{cm-cm}}$) is reported as function of time (black: raw
  data; green: moving average). Around 80 ns, a rearrangement
  of the two monomers is associated with an increase of
  $r_{\textrm{cm-cm}}$. The chain length ($r_{\textrm{end-to-end}}$)
  for the two segments (black and red for the raw data for chain A and
  B, green and blue for their moving average), are illustrated in
  Panels B. The RMSD with respect to the initial structure of the
  protein is shown in Panels C. During the time evolution, the number
  of water surrounding the protein seems not to be influenced by the
  structural changes as shown in Panels D.}
\label{fig:figmelittin} 
\end{figure}

\noindent
The structural variations together with the overall hydration of the
flexible melittin dimer in the different water box sizes are reported
in Figure \ref{fig:figmelittin}. In all simulations the end-to-end
separation of the two helices (see Figure \ref{fig:figmelittin}B) as
defined by the C$_{\alpha}$-C$_{\alpha}$ separation of the two
terminal residues Gly1 and Gln26 is stable,
indicating that the helices (the H$_2$O ratio) remain
intact. Consequently, the structural transition that occurs after 80
ns in the 51 \AA\/ box (Figure \ref{fig:figmelittin}C) and appears to
occur towards the end of the simulation in the 60 \AA\/ box involves
the dimerization interface. This is confirmed by panel A, which
reports an increase of the center-of-mass distance
$r_{\textrm{cm-cm}}$ between the two helices at the same time as the
RMSD in panel C increases. The degree of hydration (Figure
\ref{fig:figmelittin}D) defined as $r_w = N_{\rm wat} / N_{\rm max}$
remains essentially constant throughout the simulations.\\

\noindent
The protein-water interface is analysed using the same methodology as
that used for Hb. The Willard-Chandler interface is calculated setting
$r_{\rm cut}=25.0$ \AA\/ and the likelihood ($\delta
\lambda_\textrm{phob}$) of the interfacial water with the reference
TIP3P water model is determined with a 6 \AA\/ cut-off. Figures
\ref{fig:fig1} and S17 show the time evolution of selected
residues. Figure \ref{fig:fig1} shows the temporal evolution, $\delta
\lambda_\textrm{phob}$, of glycine residues Gly1, Gly3, and Gly12 for
chains A and B for both the rigid and the flexible dimers. The $\delta
\lambda_\textrm{phob}$ for the rigid dimer is essentially constant, as
expected.  The results reported are all averages over 2 ns windows.\\

\begin{table}[h]
\caption{Average $\delta \lambda_\textrm{phob}$ for the glycine
  residues for the 100 ns simulation (see Figure \ref{fig:fig1}).}
\begin{tabular}{l | r | r | r | r }
Residue   & 51 \AA\/ flexible & 51 \AA\/ rigid & 60 \AA\/ flexible & 60 \AA\/ rigid \\ 
\hline
1A  & 1.068 & 1.583 & 0.989 & 1.598 \\
3A  & 1.090 & 1.604 & 0.983 & 1.616 \\
12A & 0.965 & 0.882 & 0.979 & 0.945 \\
1B  & 1.192 & 1.614 & 0.921 & 1.587 \\
3B  & 1.156 & 1.584 & 0.927 & 1.568 \\
12B & 0.917 & 0.799 & 0.968 & 0.826 \\
\end{tabular}
 \label{tab:tab2}
\end{table}

\noindent
For the rigid monomer (blue traces in Figure \ref{fig:fig1}) the LH is
constant along the entire 100 ns simulation for both box sizes and the
averages differ by 10 \% at most (Gly12A). For the flexible dimer
(black traces) the instantaneous LH fluctuates around well-defined
average values except for Gly12B which has a slight increase in its
dynamics during the early phase of the simulation, particularly in the
60 \AA\/ box. In the simulation in both box sizes the amplitude of LH
fluctuates between 0 and 1.6, i.e. between being hydrophobic and
hydrophilic. Since Gly is an aliphatic/neutral residue, the changing
hydrophilicity must be a consequence of its embedding along the
peptide chain and the water structuring around it. Overall, it is
found that Gly12A and 12B, which are near the middle of the helix, are
less hydrophilic (see Figure \ref{fig:fig1} and Table \ref{tab:tab2})
than Gly1 and Gly3, which are positioned at or near the terminus. This
difference is more pronounced for the rigid dimer.\\

\noindent
Figure S17 shows the LH for the residues investigated by
Cheng and Rossky\cite{rossky:1998} for the 51 \AA\/ box while those
for the 60 \AA\/ box are given in Figure S18. The
average $\delta \lambda_\textrm{phob}$ are neutral or
hydrophilic. Good qualitative agreement with Ref\citep{rossky:1998} is
found for residues Val8A (hydrophilic), Leu9A, Ile13A, Ile13B
(residues with a decreasing level of hydrophobicity), and Ile20B.\\

\begin{figure}
  \centering
  \includegraphics[scale=0.50]{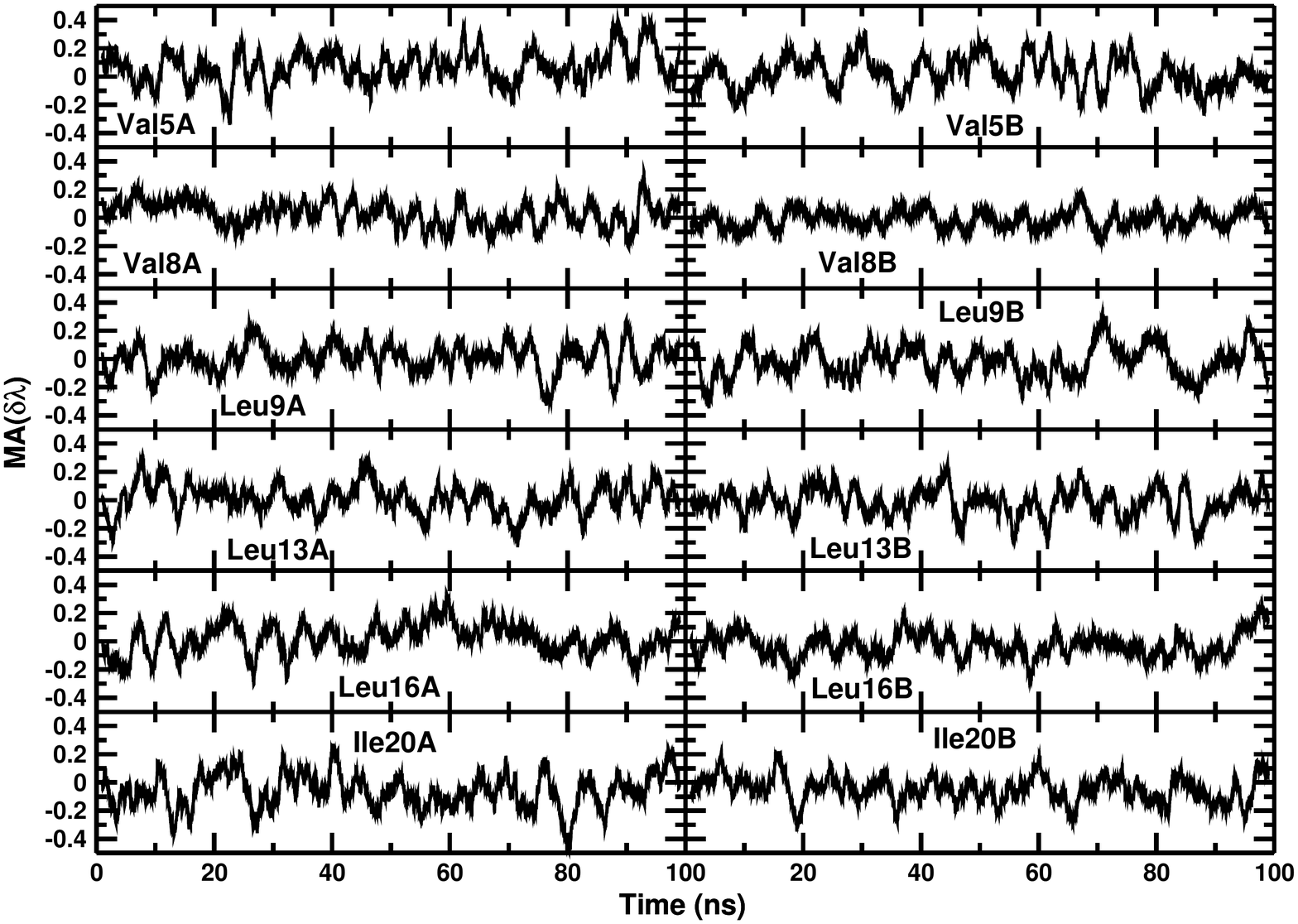}
\caption{Difference in hydrophobicity for the residues in the melittin
  dimer hydrophobic pocket as defined in Ref. \cite{rossky:1998} for
  the rigid simulations in the 51 and 60 \AA\/ boxes.  They are
  residues Val5, Val8, Leu13, Leu16, Ile20. The data reported is
  $\Delta \delta \lambda = \delta \lambda_{60} - \delta \lambda_{51}$,
  i.e. the change in LH from the simulation in the two water
  boxes. The maximum instantaneous change in $\Delta \delta \lambda
  _\textrm{phob}$ due to the box size is 40 \%; most differences are
  20 \% or less.}
\label{fig:difference} 
\end{figure}

\noindent
It is also of interest to compare the difference in hydrophobicity for
simulations of rigid melittin in the two water boxes because all
differences must arise from the size of the water box. Figure
\ref{fig:difference} shows the difference between the 51 \AA\/ and the
60 \AA\/ boxes in LH of the residues in the hydrophobic pocket.  The
average fluctuations are of the order of 0.1 units with maximum
differences of 0.4 units.  The difference between simulations with
rigid and flexible melittin can also be seen when comparing the radial
distribution functions, $g(r)$, between C$_{\alpha}$ atoms of selected
residues and water and the corresponding water occupations $N(r)$ (see
Figures S19 and S20). The residues
were chosen in accord with the results from Table \ref{tab:tab3}. For
example, in the 51 \AA\/ box for rigid melittin the values for Val5A
and Val5B are $\delta \lambda_{\rm phob} = 1.19$ and 1.48,
respectively, which change to 1.25 and 1.49 in the larger 60 \AA\/
box; i.e., this is a change of 5 \% at most.  Figure
S19A shows that $g(r)$ for Val5A and Val5B are very
similar for both water box sizes. This suggests that the difference of
$\sim 0.25$ in Table \ref{tab:tab3} for the two water boxes must arise
from the orientation of the water molecules within the 6 \AA\/
cut-off.\\

\noindent
Conversely, for flexible melittin the differences for LH in the two
water boxes can be substantial. As an example, Leu9B $\delta
\lambda_{\rm phob}^{\rm 51 A} = 0.81$ is compared with $\delta
\lambda_{\rm phob}^{\rm 60 A} =1.07$ for the two box sizes. This is
also evident from Figure S20A and B (right panel) for
which $g(r)$ and $N(r)$ have increased amplitudes for the larger water
box.  For Val5B, $\lambda_{\rm phob}^{\rm 51 A} = 1.05$ is larger in the
smaller box than
$\lambda_{\rm phob}^{\rm 60 A} = 0.86$, whereas the amplitude of $g(r)$
up to the 6 \AA\/ cut-off in the larger box is larger than that in the
smaller box. Hence, the difference found in the two box sizes must arise
from the angular orientations of the water molecules relative to the
protein surface.
These analyses suggest that both the distance-dependence (reflected
in $g(r)$ and $N(r)$) and the angular orientation, as measured by
$\lambda_{\rm phob}$, can depend on box size and potentially
influence the thermodynamic stability of the two proteins.\\

\begin{figure}
\centering
\includegraphics[scale=0.55]{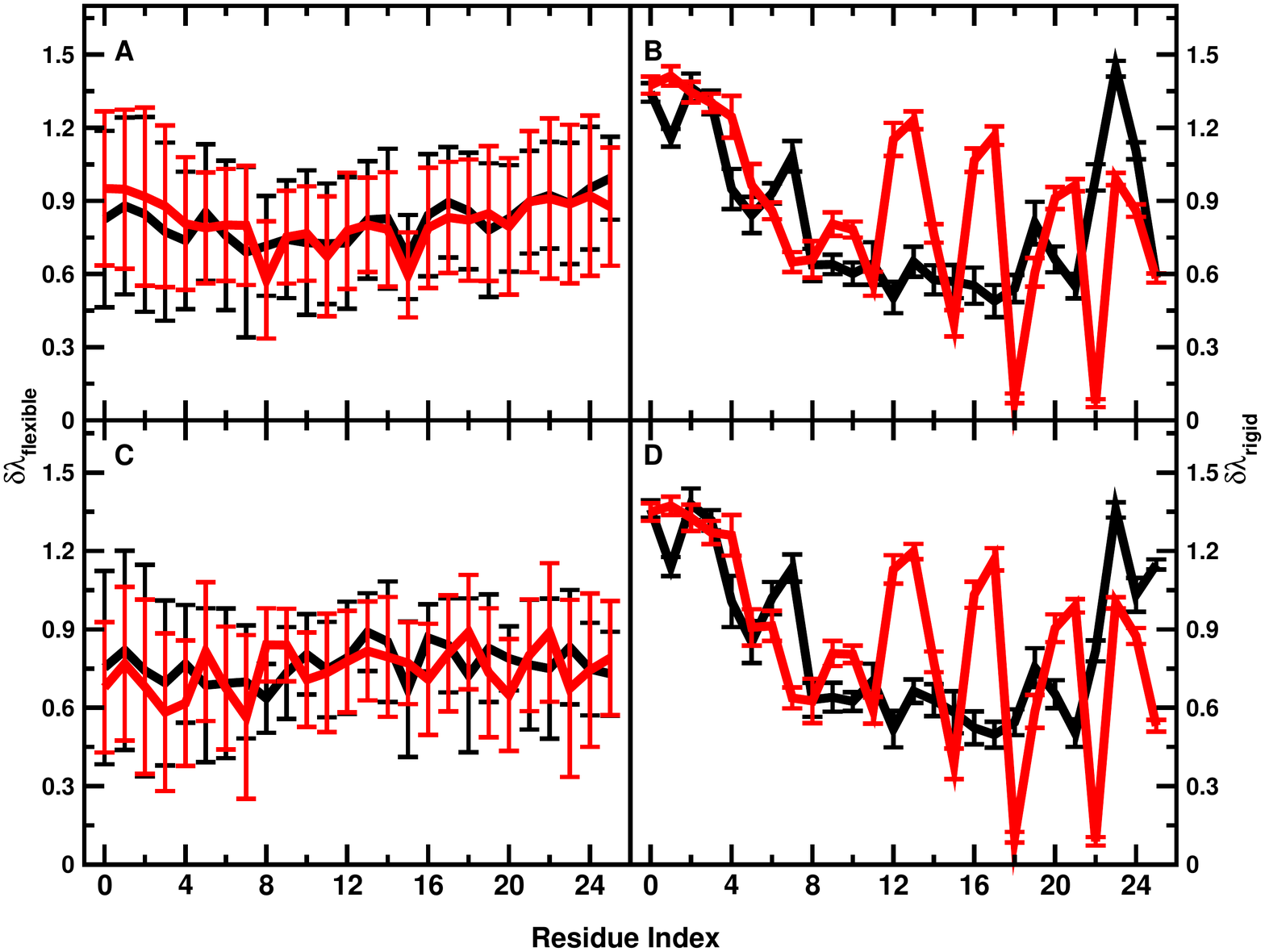} 
\caption{Average $\delta \lambda _{\rm phob}$ per residue for the 100
  ns simulation for chains A (black) and B (red) for the flexible
  (panels A and C) and the rigid (panels B and D) melittin dimer. The
  top row is for the 51 \AA\/ box and the bottom row for the 60 \AA\/
  box. The value of the LH ranges from $0.50 < \delta \lambda _{\rm
    phob} < 1.00 $ for the flexible dimer and between $0.00 < \delta
  \lambda _{\rm phob} < 1.50 $ for the rigid dimer. For the rigid
  dimer, the central part of the A-helix has consistently lower values
  of LH compared with the C- and N-terminal parts. For chain B larger
  variations in LH are found for some of the residues due to its
  different structure. The periodicity of the red traces (panels B and
  D) reflects the helical structure which is apparent for monomer B
  but less so for monomer A. For flexible melittin the variation of LH
  along the sequence is much smoother than for rigid melittin as a
  consequence of dynamical averaging.  }
\label{fig:fig14}
\end{figure}

\noindent
The average hydrophobicity for each residue during the simulation is
reported in Figure \ref{fig:fig14}A (black for chain A and red for
chain B). The decreased hydrophobicity of the central part of the
chains (from Val5 to Leu16) is highlighted by the lower $\delta
\lambda_{\rm phob}$ (of 0.1-0.2 units) compared with the C- and N-terminal
parts. The main difference between the simulation results for the
flexible (left) and rigid (right) melittin dimer is the amplitude of
the fluctuation of the hydrophobicity, but not its sign. Simulations
of rigid melittin in the 51 \AA\/ (panels A and B) and 60 \AA\/ boxes
(panels C and D) are similar to one another but they differ along the
trajectory by up to $\Delta \delta \lambda \sim 0.4$ (see Figure
\ref{fig:difference}). There are also differences between the A
(black) and B chains (red) for rigid melittin. The difference in the
monomer structures (RMSD of 1.6 \AA\/) leads to significant
differences in $\delta \lambda$ (see Figures \ref{fig:fig14}A and B.
As an example, Leu13A is considerably more hydrophobic ($\delta
\lambda_{\textrm{Leu13A}} =0.7$) than Leu13B ($\delta
\lambda_{\textrm{Leu13B}} =1.4$). Inspection of the dimer structure
shows that Leu13A points toward the dimerization interface whereas
Leu13B points away from it into the solvent. For flexible melittin
(panels A and C), the LH for the A and B chains are more similar to
one another for both box sizes. This is a consequence of
averaging along the structural dynamics.  It also suggests that
simulations on the 100 ns time scale are sufficient to converge
$\delta \lambda$ for melittin. Nevertheless, there remain certain
differences between simulations in the two water boxes for individual
residues, e.g. $\langle \delta \lambda_{\textrm{Val5B}}^{51} \rangle =
1.05$ vs. $\langle \delta \lambda_{\textrm{Val5B}}^{60} \rangle =
0.86$, see Figure \ref{fig:fig14} and Table \ref{tab:tab3}.\\

\begin{table}[h]
\caption{Average $\delta \lambda_\textrm{phob}$ for the hydrophobic
  residues from Figure 1 in Ref.\cite{rossky:1998}. The time series
  are shown in Figure S17}
\begin{tabular}{l | r | r | r | r}
Residue   & 51 \AA\/ flexible & 51 \AA\/ rigid & 60 \AA\/ flexible & 60 \AA\/ rigid\\ 
\hline
Val5A   & 0.976 & 1.185 & 1.003 & 1.246 \\                                  
Val8A   & 0.929 & 1.322 & 0.941 & 1.375 \\                                  
Leu9A   & 0.954 & 0.875 & 0.889 & 0.876 \\                                  
Leu13A  & 0.967 & 0.741 & 1.030 & 0.756 \\                                  
Leu16A  & 0.909 & 0.808 & 0.917 & 0.824 \\                                  
Ile20A  & 1.019 & 1.049 & 1.071 & 0.992 \\                                  
Val5B   & 1.046 & 1.484 & 0.861 & 1.494 \\                                  
Val8B   & 1.038 & 0.889 & 0.806 & 0.877 \\                                  
Leu9B   & 0.814 & 0.900 & 1.071 & 0.867 \\                                  
Leu13B  & 1.016 & 1.393 & 1.006 & 1.373 \\                                  
Leu16B  & 0.835 & 0.637 & 1.012 & 0.626 \\                                  
Ile20B  & 1.087 & 0.845 & 0.988 & 0.824 \\                                  
\end{tabular}
 \label{tab:tab3}
\end{table}

\noindent
Finally, the two-dimensional LH-CCM (see Figure S21) for
all four systems and their differences between rigid and flexible
dimer (bottom row of Figure S21) have been determined. As
can be anticipated from Figure \ref{fig:difference}, the LH cross
correlation maps for the two box sizes are similar for rigid melittin
dimer (see Figure S22 left panel). On the other hand,
the differences between rigid and flexible melittin dimer in the two
boxes are considerably larger, as the bottom row of Figure
S21 demonstrates. While for the 51 \AA\/ box differences
primarily occur at the interface (upper left quadrant), differences
for the larger 60 \AA\/ box occur both at the interface and along the
two helices. Furthermore, the amplitude of the differences increases
in going from the 51 \AA\/ to the 60 \AA\/ box.\\

\noindent
The difference between flexible melittin dimer in the 51
\AA\/ and 60 \AA\/ boxes is shown in Figure S22,
right panel. Increasing the box size leads to more pronounced
cross correlation peaks between the beginning of helix A and the end
of helix B. A slightly less pronounced increase in the correlation is
found for the end of chain A and the beginning of chain
B. Corresponding radial distribution functions are reported in Figure
S23. Even for rigid melittin in the 51 \AA\/ (red) and
60 \AA\/ (blue) boxes (e.g. for Gly3A and Gln26B) there are slight
differences between the $g(r)$. Compared with rigid melittin, the
$g(r)$ for flexible melittin are all less structured. Except for Ile2B
they also agree well for the two box sizes.\\

\noindent
For the rigid and flexible dimer, differences for the two box sizes
also occur as shown in the $\Delta$CCC map (Figure
S22). They effect the local hydrophobicity which is
computed from the water structuring. For the rigid protein surface the
differences are essentially independent of box size, as shown in the
left panel. For the flexible dimer there are significant differences.
They arise both from protein structural changes and the surrounding
water structuring. In this context it is interesting to note that
$g(r)$ around Leu9A and Leu9B in Figure S20 in the 60
\AA\/ box are virtually identical, whereas the LH differs by almost 15
\% (0.89 vs. 1.07, see Table \ref{tab:tab3}). As LH includes both the
distance between the water oxygen atom relative to the protein surface
and the angular orientation of the OH vector, this difference in LH is
likely to be related to different orientations of the water network
around Leu9A and Leu9B.\\

\section{Conclusions}
The present work analyzed the local hydrophobicity around key residues
at the protein interfaces for hemoglobin and melittin. It was found
that the local hydrophobicity measure for Hb provides valuable insight
into the effect of different box sizes from MD
simulations. Specifically, analysis of the local hydrophobicity cross
correlation coefficients for Hb provided a dynamical view of Perutz's
stereochemical model involving breaking and formation of salt bridges
at the $\alpha_1 / \beta_2$ and $\alpha_2 / \beta_1$ interfaces. Also,
the more detailed analysis of the simulations in the 90 \AA\/ and 120
\AA\/ boxes demonstrates that they decay to known but different
intermediate structures upon destabilization of the $\alpha / \beta$
interface following a decrease in LH, i.e. as a consequence of reduced
water density or change of water orientation at the protein/water
interface. This is consistent with earlier findings\cite{MM.hb:2018}
that reported a reduced number of water-water hydrogen bonds for the
smaller boxes, which influences the equilibrium between water-water
and water-protein contacts and hence the water activity. The present
results also support recent extensive simulation studies of the
A$\beta$ peptide which show that the hydrophobic surface area
increases significantly in small cells along with the standard
deviation in exposure and backbone conformations.\cite{kepp:2019} As
is also reported here (see Figure \ref{fig:shinv2}), hydrophilic
exposure was found to dominate in large boxes whereas hydrophobic
exposure is prevalent in small cells.  This suggests there is a
weakening of the hydrophobic effect in smaller water box sizes.\\

\noindent
Early experiments indicate that T$_0$ is significantly ($\sim 8$
kcal/mol, equivalent to $K_{\frac{T_0}{R_0}}=6.7 \times 10^5$) more
stable than R$_0$.\cite{edelstein:1971} Also, the rate for the R$_0$
$\rightarrow$ T$_0$ has been determined as $15700 \pm 700$ s$^{-1}$ at
303 K, corresponding to a transition time of $\sim 20$
$\mu$s.\cite{sawicki:1976} As shown in Ref. \cite{cui:2008} this
implies that the T$_0$$\rightarrow$R$_0$ transition occurs on a time
scale of 1 to 10 s by use of the Arrhenius equation. This is far too
long to be sampled directly by MD simulations with explicit solvent in
a statistically meaningful way. As an example, for association free
energies in protein-ligand and protein-protein interactions from
replica exchange coarse grained simulations, a total simulation time
of $> 5$ $\mu$s was deemed necessary for convergence\cite{sansom:2017}
and similar studies were carried out for protein-ligand interactions
using atomistic force fields.\cite{MM.stab:2018,vu:2019} Alternative
approaches, such as conjugate peak refinement, string methods, or
nudged elastic band in explicit solvent are also ways to more
quantitatively investigate the transition state
region.\cite{karplus.hb:2011,ovchi:2011,jonsson:2000} However, to
quantify differences between the T$_0$ and R$_0$ structure or
structures evolving towards the R$_0$ state (as done here), explicit
knowledge of the transition state region is not required.\\

\noindent
For the melittin dimer the role of box size on the hydration dynamics
was expected to be smaller, based on earlier work on the hydrophobic
effect.\cite{chandler:2005} Nevertheless, the analysis of rigid
melittin dimer, which was studied in previous work,\cite{rossky:1998}
suggests that the water distribution is affected by the box size for
the 51 \AA\/ and 60 \AA\/ boxes.  These differences become more
pronounced when the protein structure is allowed to change in the
simulations.\\

\noindent
Complementary to radial distribution functions $g(r)$, the local
hydrophobicity (LH) provides a time-dependent quantitative local measure
characterizing the water dynamics and structure
around a protein. When combined with time-dependent structural
information a more complete picture for the coupled protein-water
dynamics emerges. It provides valuable information about
thermodynamic manifestations of structural changes at a molecular
level.\\

\section*{Data and Code Availability}
The water-structure analysis code used to calculate
$\delta\lambda_{phob}$ is publicly available at
\url{https://github.com/mjmn/interfacial-water-structure-code}.

\section*{Acknowledgment}
Support by the Swiss National Science Foundation through grants
200021-117810, the NCCR MUST (to MM), and the University of Basel is
acknowledged. The support of MK by the CHARMM Development Project is
gratefully acknowledged. AW, MN, and SS were supported by the National
Science Foundation under CHE-1654415\\

\bibliography{article}
\end{document}